\documentclass[a4paper,aps,11pt,nofootinbib,pdftex]{revtex4}
\usepackage{amsmath}
\usepackage{amssymb}
\usepackage{amsfonts}
\usepackage{color}
\usepackage{float}
\usepackage{comment}
\usepackage{ulem}
\usepackage{ascmac}
\usepackage{setspace}

\usepackage{graphicx}

\begin{document}

\def\uwave{ }
\title{\Large
A Variety of Nontopological Solitons \\
in a Spontaneously Broken U(1) Gauge Theory
\\
\large\sl - Dust Balls, Shell Balls, and Potential Balls -
}

\begin{spacing}{1.0}
\hfill{OCU-PHYS 532}

\hfill{AP-GR 166}

\hfill{NITEP 95}
\end{spacing}

\author{\vspace{1cm}
\Large Hideki Ishihara}
\email{ishihara@sci.osaka-cu.ac.jp}
\author{\Large Tatsuya Ogawa}
\email{taogawa@sci.osaka-cu.ac.jp}
\affiliation{
 Department of Mathematics and Physics,
 Graduate School of Science,
Nambu Yoichiro Institute of Theoretical and Experimental Physics (NITEP),
Osaka City University, Osaka 558-8585, Japan}

\begin{abstract}
\vspace{0.5cm}
We show, by numerical calculations, that there exist three-types of stationary and
spherically symmetric nontopological soliton solutions (NTS-balls) with large sizes
in the coupled system consisting of a complex matter scalar field, a U(1) gauge field,
and a complex Higgs scalar field that causes spontaneously symmetry breaking.
Under the assumption of symmetries, the coupled system reduces
to a dynamical system with three degrees of freedoms governed by an effective action.
The effective potential in the action has stationary points.
The NTS-balls with large sizes are described by bounce solutions that
start off an initial stationary point, and traverse to the final stationary point,
vacuum stationary point.
According to the choice of the initial stationary point,
there appear three types of NTS-balls: dust balls, shell balls, and potential balls
with respect to their internal structures.
\end{abstract}

\maketitle

\section{Introduction}

A nontopological soliton (NTS) is a localized configuration of fields
whose stability is guaranteed by a conserved Noether charge,
and it can be interpreted as a condensation of bosonic particles in a bound state.
In the pioneering work by Friedberg, Lee, and Sirlin \cite{Friedberg:1976me},
the NTS solutions are constructed in a field theory that consists of a complex scalar field
and a real scalar field with a double-well potential that causes
the spontaneous symmetry breaking.
Coleman \cite{Coleman:1985ki} constructed simplest NTSs, so-called \textit{Q-balls},
in a theory of a single complex scalar field with a complicated self-interaction.

Extension of the NTS solutions have been investigated in cosmology and astrophysics.
There are lots of works in the context that the NTS
is a candidate of dark matter \cite{Kusenko:1997si, Kusenko:2001vu,
Fujii:2001xp,Enqvist:2001jd, Kusenko:2004yw}
and a source for a baryogenesis \cite{Enqvist:1997si,Kasuya:1999wu,Kawasaki:2002hq}
in a wide class of field theories with a potential
inspired by a supersymmetric theories \cite{Kusenko:1997zq,Dvali:1997qv,Kasuya:2000sc}.

As natural extensions of the NTSs in a theory with a gauge field were
studied in the works \cite{Lee:1988ag,Shi:1991gh, Gulamov:2015fya}.
In their papers, it was clarified that gauged NTSs with a large amount of charge are unstable
because of the Coulomb repulsive force mediated by the gauge field.
Then, the gauged NTSs have an upper limit of the charge.
It is also interesting issue to investigate NTSs in spontaneously broken gauge theories,
which is a standard framework of the gauge theories.
It was shown that stable gauged NTS solutions with a large amount of charge exist
in a spontaneously broken gauge
theory \cite{Ishihara:2018rxg,Ishihara:2019gim,Forgacs:2020vcy}.

In this paper, we investigate the NTS solutions in the system
that consists of a matter complex scalar field, a U(1) gauge field,
and a complex Higgs scalar field that causes the spontaneous symmetry breaking.
Under the assumption that the solutions are stationary and spherically symmetric,
the system reduces to a theory described by an effective action
with three dynamical degrees of freedom.
The effective potential of the theory has stationary points, and there exist
solutions that connect the stationary points as bounce solutions.
The bounce solutions represent NTS balls with large size.
The solutions are classified into three types:
balls filled by homogeneous dust inside (dust balls),
empty balls with shells at their surfaces (shell balls),
balls filled by potential energy of the Higgs scalar field in the surface shells (potential balls).
In all of NTS balls, U(1) charge carried by the matter scalar field is compensated
by the counter charge carried by the Higgs scalar field.
Namely, all NTS balls are totally screened \cite{Ishihara:2018rxg,Ishihara:2019gim,Forgacs:2020vcy}.

The paper is organized as follows.
In Section II, we present the basic model studied in this article, and show that
the system is reduced to a dynamical system with three degrees of freedom.
In Section III, we show that the effective potential of the system has stationary points,
and the bounce solutions that connect two of the stationary points represent
NTS balls with large sizes.
In Section IV, we obtain three types of the bounce solutions numerically,
and study dynamical properties of the solutions.
In Section V, we study the internal properties of the NTS ball solutions,
and in Section VI, we discuss stability of the balls briefly.
Section VII is devoted to summary.

\newpage

\section{Basic Model}

The action of the system we consider is
\begin{align}
	S=\int d^4 x \left(
  		-\left( D_{\mu}\psi \right)^{\ast}\left( D^{\mu}\psi \right)
		-\left(D_{\mu}\phi \right)^{\ast}\left(D^{\mu}\phi \right)
		-V(\phi)
		-\mu \psi^{\ast}\psi \phi^{\ast}\phi
		-\frac{1}{4}F_{\mu\nu}F^{\mu\nu}
		\right),
\label{eq:action}
\end{align}
where $\psi$ is a complex matter scalar field, $\phi$ is a complex Higgs scalar field
with the potential
\begin{align}
		V(\phi) = \frac{\lambda}{4}(\phi^{\ast}\phi-\eta^2)^2
\end{align}
characterized by constsnts $\lambda$ and $\eta$,
and $F_{\mu\nu}:=\partial_{\mu}A_{\nu}-\partial_{\nu}A_{\mu}$ is the field strength of
a U(1) gauge field $A_{\mu}$.
The parameters $\mu$ is the coupling constant between $\psi$ and $\phi$, and the covariant
derivative $D_{\mu}$ in \eqref{eq:action} is defined by
\begin{align}
  	D_{\mu}\psi =\partial_{\mu}\psi -ieA_{\mu}\psi,  \quad
	D_{\mu}\phi =\partial_{\mu}\phi -ieA_{\mu}\phi,
 \label{eq:covariant_derivative}
\end{align}
where $e$ is a gauge coupling constant.

The system is invariant under
the $\text{U}_\text{local}(1) \times \text{U}_\text{global}(1)$ transformations:
\begin{align}
  	&\psi(x) \to \psi'(x)=e^{i(\chi(x)-\gamma)}\psi(x),
 \label{eq:psi_tr} \\
  	&\phi(x) \to \phi'(x)=e^{i(\chi(x)+\gamma)}\phi(x),
 \label{eq:phi_tr} \\
 	&A_{\mu}(x)\to A_{\mu}'(x)=A_{\mu}(x)+e^{-1}\partial_{\mu}\chi(x),
 \label{eq:A_tr}
\end{align}
where $\chi(x)$ is an arbitrary function, and $\gamma$ is a constant.
The conserved currents associated with the invariance are
\begin{align}
  j_\psi^{\nu} =ie\left(\psi^{\ast}(D^{\nu}\psi)-\psi(D^{\nu}\psi)^{\ast}\right), ~~
  j_\phi^{\nu} =ie\left(\phi^{\ast}(D^{\nu}\phi)-\phi(D^{\nu}\phi)^{\ast}\right),
  \label{eq:j_psi_phi}
\end{align}
which satisfy $\partial_{\mu}j_{\psi}^{\mu}$=0 and $\partial_{\mu}j_{\phi}^{\mu}$=0,
and the total conserved charge of $\psi$ and $\phi$ are defined by volume integrations
on a $t=const.$ plane $\Sigma$ in the form
\begin{align}
  Q_\psi =\int \rho_{\psi} ~dV, ~~ Q_\phi :=\int \rho_{\phi} ~dV,
  \label{eq:Q_psi_phi}
\end{align}
where $\rho_{\psi}:=j_{\psi}^t$ and $\rho_{\phi}:=j_{\phi}^t$.

In the symmetry breaking vacuum where
the Higgs scalar field has the vacuum expectation value $\langle \phi \rangle=\eta$,
the gauge field $A_{\mu}$ and the complex matter scalar field $\psi$ acquire
the mass $m_A=\sqrt{2}e\eta$ and $m_{\psi}=\sqrt{\mu}\eta$, respectively.
Simultaneously, the real part
of $\phi$ around $\eta$ acquires the mass $m_\phi=\sqrt{\lambda}\eta$.
From the action (\ref{eq:action}),
we obtain the Klein-Gordon equations and the Maxwell equation in the form
\begin{align}
 &D_{\mu}D^{\mu}\psi-\mu |\phi|^2 \psi =0,
  \label{eq:equation of psi}\\
 &D_{\mu}D^{\mu}\phi-\frac{\lambda}{2}\phi(|\phi|^2-\eta^2)-\mu \phi |\psi|^2 =0,
  \label{eq:equation of phi}\\
 &\partial_{\mu}F^{\mu\nu}=j_\phi^{\nu}+j_{\psi}^{\nu}.
  \label{eq:equation of gauge field}
\end{align}

In order to construct stationary and spherically symmetric solutions,
we use the following ansatz:
\begin{align}
  	\psi=e^{i\omega t}u(r), \quad \phi=e^{i\omega' t}f(r), \quad A_t=A_t(r),\quad
	\mbox{and}\quad A_i=0 \quad \text{for}~i=r,\theta,\varphi ,
  \label{eq:ansatz}
\end{align}
in the spherical coordinates $(t, r, \theta, \varphi)$,
where $\omega$ and $\omega'$ are constants, and $u(r)$ and $f(r)$ are real functions of $r$.
Using the gauge transformation \eqref{eq:psi_tr}, \eqref{eq:phi_tr} and \eqref{eq:A_tr},
we fix the variables as
\begin{align}
  &\phi(r) \to f(r),
\label{eq:f}\\
  &\psi(t,r) \to e^{i\Omega t}u(r)
\label{eq:u}\\
  &A_t(r) \to \alpha(r):= A_t(r)-e^{-1}\omega',
\label{eq:alpha}
\end{align}
where $\Omega:=\omega-\omega'$, and we assume it positive without loss of generality.
Therefore, we rewrite
\eqref{eq:equation of psi}, \eqref{eq:equation of phi}, and \eqref{eq:equation of gauge field},
as the coupled system of nonlinear ordinary differential equations in the form
\begin{align}
 &\frac{d^2u}{dr^2}+\frac{2}{r}\frac{du}{dr}+(e\alpha-\Omega)^2u-\mu f^2u=0,
 \label{eq:eq_u}\\
 &\frac{d^2f}{dr^2}+\frac{2}{r}\frac{df}{dr}+e^2f \alpha^2-\frac{\lambda}{2}f(f^2-\eta^2)-\mu fu^2=0,
\label{eq:eq_f}\\
 &\frac{d^2\alpha}{dr^2}+\frac{2}{r}\frac{d\alpha}{dr}+\rho_\text{total}=0,
\label{eq:eq_alpha}
\end{align}
where $\rho_\text{total}(r):=\rho_{\psi}(r)+ \rho_{\phi}(r)$, and
$\rho_\psi$ and $\rho_\phi$ are given by
\begin{align}
  \rho_\psi = -2e(e\alpha-\Omega) u^2,
  \quad
	\rho_\phi = -2e^2\alpha f^2.
\label{eq:rho_psi_phi}
\end{align}

We require that the fields are regular at the origin of spherical symmetry,
and are in the vacuum at the infinity.
Then, we impose the conditions
\begin{align}
  \frac{du}{dr}\to 0 \ , \
  \frac{df}{dr}\to 0 \ , \ \frac{d\alpha}{dr}\to 0 \quad \mbox{as}\quad r\to 0,
\label{eq:BC_origin}
\end{align}
and
\begin{align}
  u \to 0 \ , \ f \to \eta \ , \ \alpha \to 0 \quad \mbox{as}\quad r\to \infty .
\label{eq:BC_infty}
\end{align}

\section{Stationary Points of the System and Bounce Solutions}

\def\particle{\lq particle\rq\ }

The coupled system of the equations \eqref{eq:eq_u}, \eqref{eq:eq_f}, and \eqref{eq:eq_alpha}
can be derived from the effective action
\begin{align}
	&S_\text{eff} = \int  r^2 dr \left(
		 \biggl(\frac{du}{dr}\biggr)^2 + \biggl(\frac{df}{dr}\biggr)^2
		-\frac12 \biggl(\frac{d\alpha}{dr}\biggr)^2
		- U_\text{eff}(u, f,\alpha)	\right),
\label{eq:S_eff}
\end{align}
where $U_\text{eff}$ is given by
\begin{align}
	&U_\text{eff}(u, f,\alpha) = -\frac{\lambda}{4}(f^2-\eta^2)^2-\mu f^2 u^2
		+e^2f^2 \alpha^2 +(e\alpha-\Omega)^2u^2 .
\label{eq:U_eff}
\end{align}
Regarding the radius $r$ as a \lq time\rq \  and amplitude of the fields $u(r), f(r)$,
and $\alpha(r)$ as the position of a \particle, we can understand the effective action
\eqref{eq:S_eff} in the words of Newtonian mechanics of three degrees of freedom.
We note two points:
the first, the \lq kinetic\rq \ term of $\alpha$ has the wrong sign;
the second, $r$ appears explicitly in the integration measure
in the effective action \eqref{eq:S_eff}, so that the equations of motions
\eqref{eq:eq_u}-\eqref{eq:eq_alpha} have friction terms that are proportional to $1/r$.

The dynamical system described by the effective action has stationary points of $U_\text{eff}$
that satisfy
\begin{align}
	\frac{\partial U_\text{eff}}{\partial u}=0,
\quad
	\frac{\partial U_\text{eff}}{\partial f}=0,
\quad \text{and}\quad
	\frac{\partial U_\text{eff}}{\partial \alpha}=0.
\label{eq:stationary_cond}
\end{align}
By solving the coupled equations \eqref{eq:stationary_cond}, we find two isolated
stationary points, $\rm P_v$ and $\rm P_0$,
and two ridges, $\rm R_1$ and $\rm R_2$, each ridge consists of infinite stationary
points aligned on a line.
The positions of the stationary points
in the region $u\geq 0, f\geq 0$, and $\alpha\geq 0$
are
\begin{align}
    {\rm P_v} \ : \
  	&\alpha_\text{v} =0, \quad f_\text{v} =\eta, \quad u_\text{v} =0,
    \label{eq:position_Pv}
\\
  \rm P_0 \ : \
	&\alpha_0 =\frac{1}{e(4\mu-\lambda)}
		\left((\mu-\lambda)\Omega+\sqrt{\mu(2\lambda+\mu)\Omega^2
		-\mu\lambda(4\mu-\lambda)\eta^2}\right),
\cr
	&f_0 = \frac{1}{\sqrt{\mu}}(\Omega -e\alpha_0),
 	\quad u_0 = \frac{1}{\sqrt{\mu}}\sqrt{e\alpha_0(\Omega -e\alpha_0)},
  \label{eq:position_P0}
\\
  {\rm R_1} \ : \
	&\alpha_1 =\frac{\Omega}{e}, \quad f_1 =0, \quad u_1 =\text{arbitrarly constants},
  \label{eq:position_P1}
\\
  {\rm R_2}  \ : \
	&\alpha_2 =\text{arbitrarly constants}, \quad f_2 =0, \quad u_2 =0.
  \label{eq:position_P2}
\end{align}
The point ${\rm P_v}$ is the vacuum  stationary point, i.e., the conditions \eqref{eq:BC_infty}
are satisfied there.
There are copies of the stationary points in the regions of possible alternative signs of $u$, $f$,
and $\alpha$, respectively.

We concentrate on solutions that connect one of the stationary points
and the vacuum stationary point, namely, at the initial \lq time\rq, $r=0$,
a \particle in the space $(u, f, \alpha)$
stays in a vicinity of one of the stationary points long \lq time\rq .
The \particle leaves the stationary point and traverses quickly toward the vacuum stationary point,
$\rm P_v$, and finally stays on it.
We call the solutions that connect one of the stationary points and the vacuum stationary
point {\sl bounce solutions}.

For the bounce solutions, the equations of motion have at least one unstable direction on both the initial stationary point and the vacuum stationary point.
As is shown in Appendix \ref{stability}, there are unstable directions at the stationary points
$\rm P_v, P_0$, and any points $\rm P_1$ on $\rm R_1$, while there is no unstable direction
at any point on $\rm R_2$.
Then, possible bounce solutions are in the following three types:
\begin{align}
	\rm P_0 \to P_v, \quad
	\rm P_1 \to P_v, \quad
	\rm P_v \to P_v.
\label{bound_sol}
\end{align}
Hereafter, these three types are denoted by (0-V), (1-V), and (V-V), respectively.

Insert $(u, f, \alpha)$ of the stationary points $\rm P_v, P_0, P_1$
given by \eqref{eq:position_Pv}, \eqref{eq:position_P0}, and \eqref{eq:position_P1} into \eqref{eq:U_eff},
we obtain the values of $U_\text{eff}$ at the stationary points as
\begin{align}
	U_\text{eff}({\rm P_v})=0, \quad U_\text{eff}(\rm P_1)=-\lambda \eta^4/4,
\end{align}
and $U_\text{eff}(\rm P_0)$ is given by a complicated function of the parameter $\Omega$.
It should be noted that in the limit
\begin{align}
  \Omega\to \Omega_\text{min}:= \sqrt{2\sqrt{\lambda\mu}-\lambda}~ \eta
		=\sqrt{m_{\phi}(2m_{\psi}-m_{\phi})}
 \label{eq:Omega_limit_min}
\end{align}
we see
\begin{align}
	U_\text{eff}(\rm P_0) \to 0.
\end{align}

We can define the effective energy of solutions as the sum of the kinetic energy and
the effective potential in the form:
\begin{align}
	E_\text{eff}=\biggl(\frac{du}{dr}\biggr)^2
		+ \biggl(\frac{df}{dr}\biggr)^2 -\frac12 \biggl(\frac{d\alpha}{dr}\biggr)^2
		+ U_\text{eff}(u,f,\alpha).
\label{eq:E_eff}
\end{align}
We can also define the work done by the friction force acting on a moving \particle
in the form:
\begin{align}
	W =-2\int \frac{2}{r}\left(\frac{du}{dr}\right)du
		-2\int \frac{2}{r}\left(\frac{df}{dr}\right)df
		+\int \frac{2}{r}\left(\frac{d\alpha}{dr}\right)d\alpha.
  \label{eq:E_fri}
\end{align}
The value of the effective energy changes by an equal amount of $W$.
Note that the equations of motion of $f$ and $u$ have friction terms,
while the equation of motion of $\alpha$ has an anti-friction term in \eqref{eq:E_fri}
since the kinetic term of the gauge field has the wrong sign.
Therefore, as the \lq time\rq \ is increasing, the motions of scalar fields cause a decrease of $E_\text{eff}$,
while the motion of gauge field causes an increases of $E_\text{eff}$.

\section{Numerical calculations}

In this section, we present numerical solutions of NTS-balls described
by the bounce solutions for the coupled equations
of motion \eqref{eq:eq_u}, \eqref{eq:eq_f}, \eqref{eq:eq_alpha}.
In order to search solutions that satisfy the boundary conditions \eqref{eq:BC_origin}
and \eqref{eq:BC_infty},
we should tune the parameter $\Omega$ and boundary values of the fields.
To achieve this, we use the relaxation method.
In numerics, hereafter, we set $\eta$ as the unit.
We set $\lambda=1$ and $\mu=1.4$, as an example,
and we consider two cases, $e=1.0$ and $e=0.10$.

\subsection{Field configurations}
\begin{figure}[H]
\centering
\includegraphics[width=5.5cm]{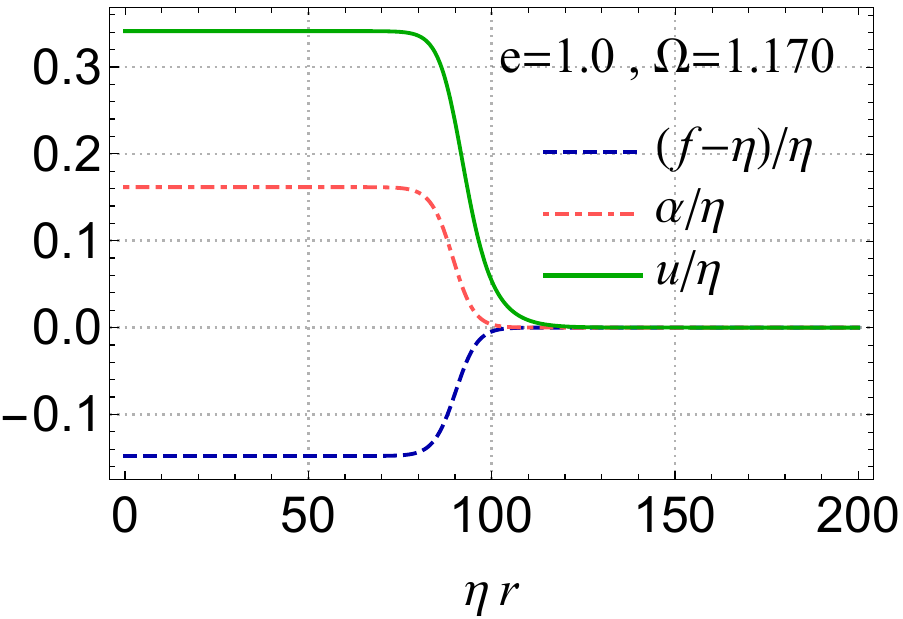}~~~~
\includegraphics[width=5.2cm]{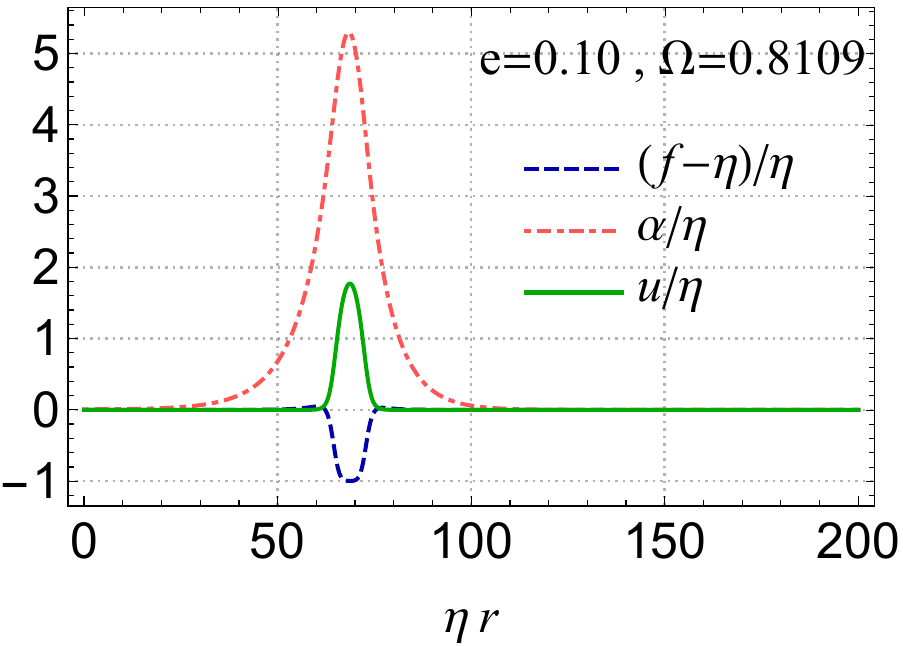}~~~~
\includegraphics[width=5.2cm]{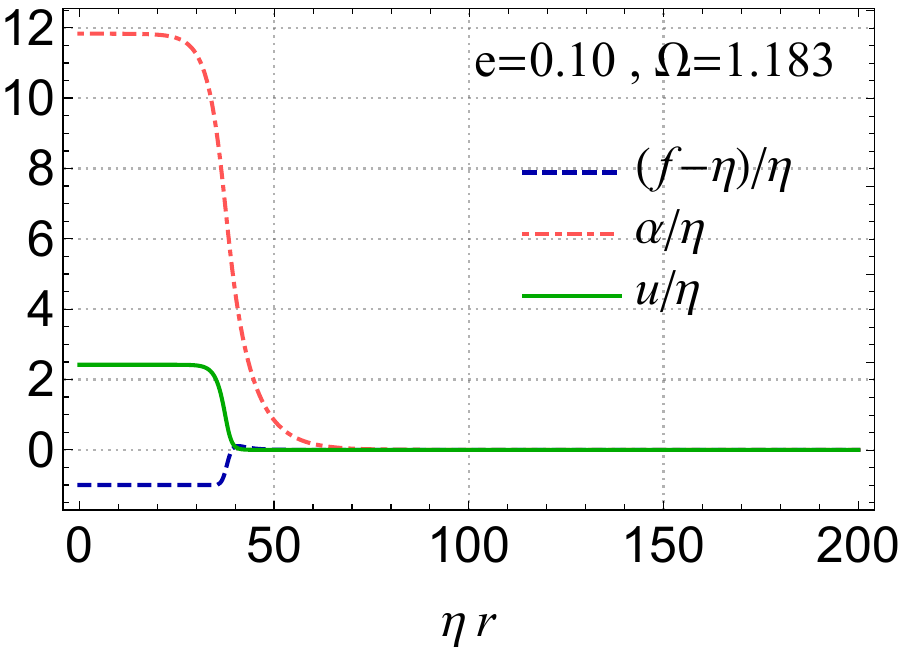}
\caption{
Typical numerical solutions, $f$, $u$, and $\alpha$ are shown as functions of $r$
for (0-V) type (left panel), (V-V) type (central panel), and (1-V) type (right panel).
\label{fig:configurations}
}
\end{figure}

In Fig.\ref{fig:configurations}, we plot typical configurations of $u$, $f$, and $\alpha$
as functions of $r$ obtained by numerical calculations.
In the case of $e=1.0$, only (0-V) type solutions (left panel of Fig.\ref{fig:configurations})
appear as bounce solutions~\cite{Ishihara:2018rxg,Ishihara:2019gim,Forgacs:2020vcy}.
On the other hand, in the case of $e=0.10$, two types, (V-V) (central panel)
and (1-V) (right panel), appear.

For solutions of two types, (0-V) and (1-V), fields behave like step functions,
namely, the functions $u,f$ and $\alpha$ take constant values inside
a characteristic radius, say $R$, around which
the functions decay quickly to the vacuum values.
These solutions represent homogeneous balls with the surface radius $R$.
The central values of $u,f$, and $\alpha$ of (0-V) type solutions are given
by \eqref{eq:position_P0},
while for (1-V) type solutions, $f$ and $\alpha$ are given by \eqref{eq:position_P1},
and the value of $u$, i.e.,
the position of the initial stationary point on the ridge $\rm R_1$, is determined
by global behavior of the solutions.
In the case of (V-V) type solutions, the functions $u,f$, and $\alpha$ take
non-vacuum values only in a vicinity of a characteristic radius,
namely, the solutions represent empty balls surrounded by spherical shells
with this radius\footnote{Shell-like solutions were found in a system
consists of a complex scalar field with a singular potential coupled to
a U(1) gauge field~\cite{Arodz:2008nm,Tamaki:2014oha}.}.

\begin{figure}[h]
\centering
\includegraphics[width=12.0cm]{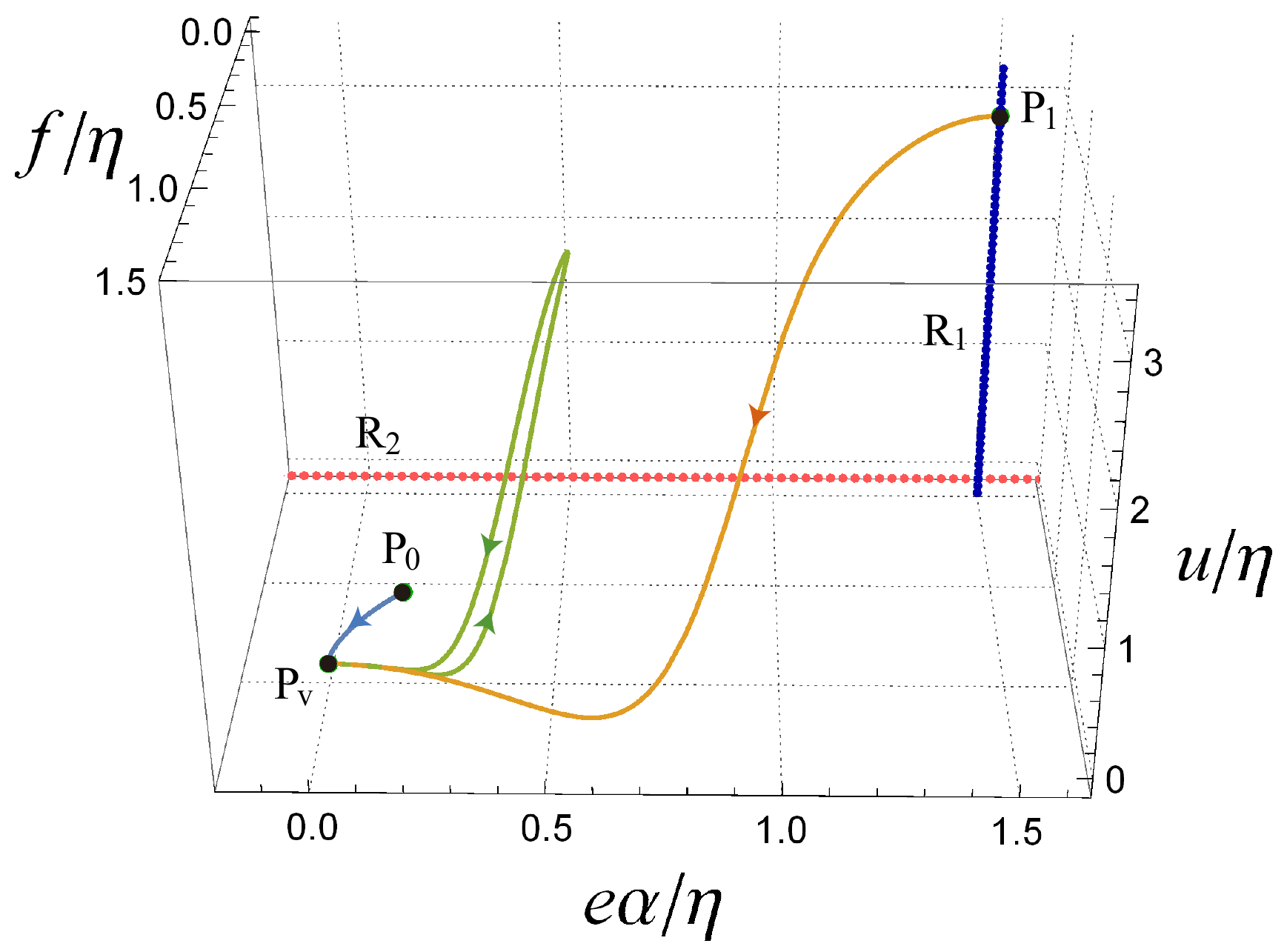}~~~~
\caption{
Trajectories of the moving \particle that connects the stationary points:
$\rm P_0 \to P_v$ (blue),  ${\rm P_v \to P_v}$ (green), and ${\rm P_1 \to P_v}$ (orange).
The doted lines $\rm R_1$ and  $\rm R_2$ are the ridges that consist of infinite numbers of
stationary points.
\label{fig:trajectory_bounce_solution}
}
\end{figure}

In Fig.\ref{fig:trajectory_bounce_solution}, we show the positions of stationary points,
${\rm P_0}$, ${\rm P_v}$, and ridges, ${\rm R_1}$, ${\rm R_2}$, of the effective potential
$U_\text{eff}$ in the $(u,f,\alpha)$ space.
Global behavior of trajectories of moving \lq particles\rq \ in the $(u,f,\alpha)$ space
that describe the bounce solutions are shown in the same figure.
Actually, the trajectories connect a stationary point and the vacuum stationary point.

\begin{figure}[H]
\centering
\includegraphics[width=5.5cm]{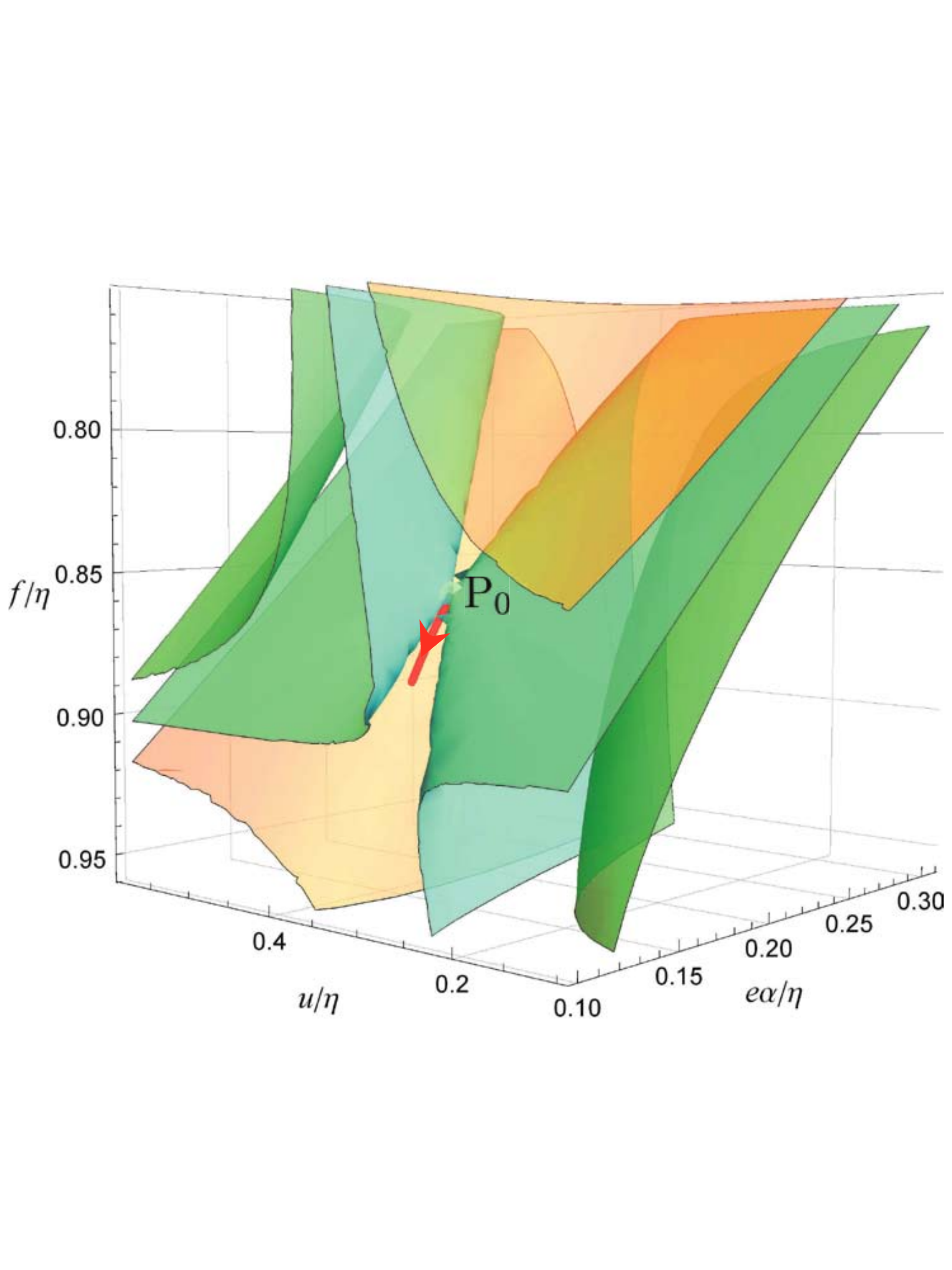}
\hspace{15mm}
\includegraphics[width=5.5cm]{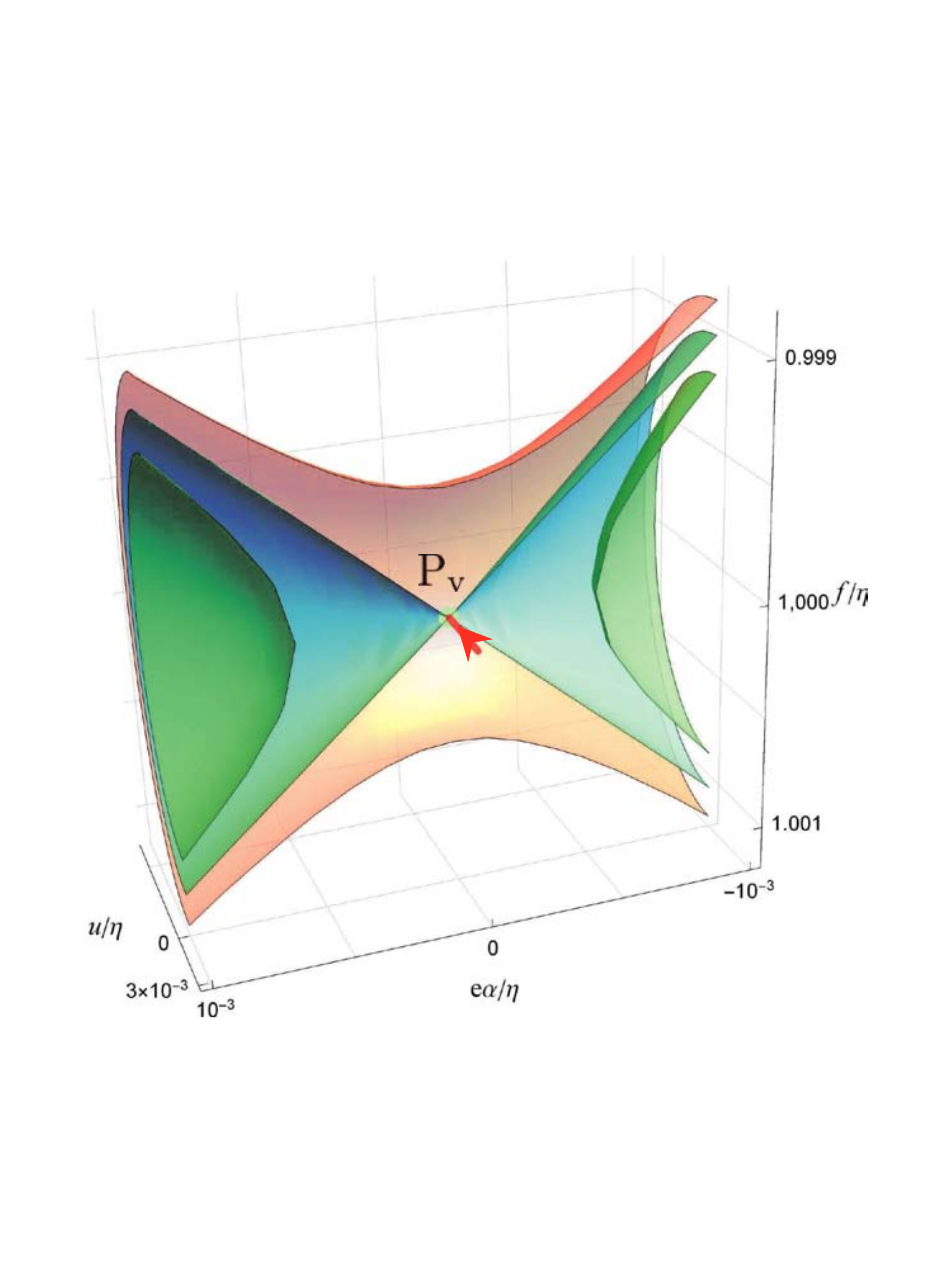}
\\
\vspace{-5mm}
(i) ${\rm P_0} \to {\rm P_v} \ : \ (e=1.0 \ , \ \Omega=1.170)$
\\
\includegraphics[width=5.5cm]{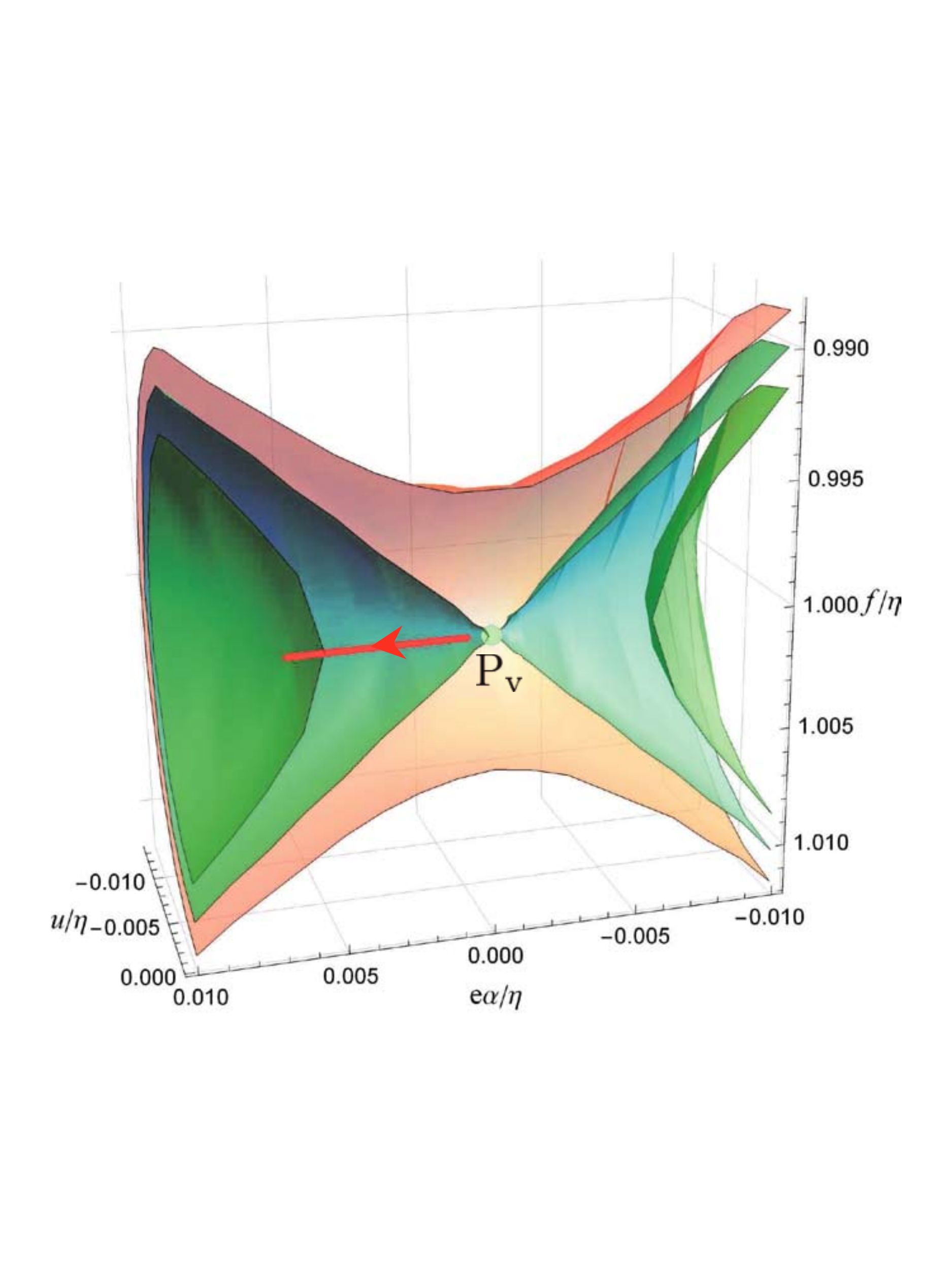}
\hspace{15mm}
\includegraphics[width=5.5cm]{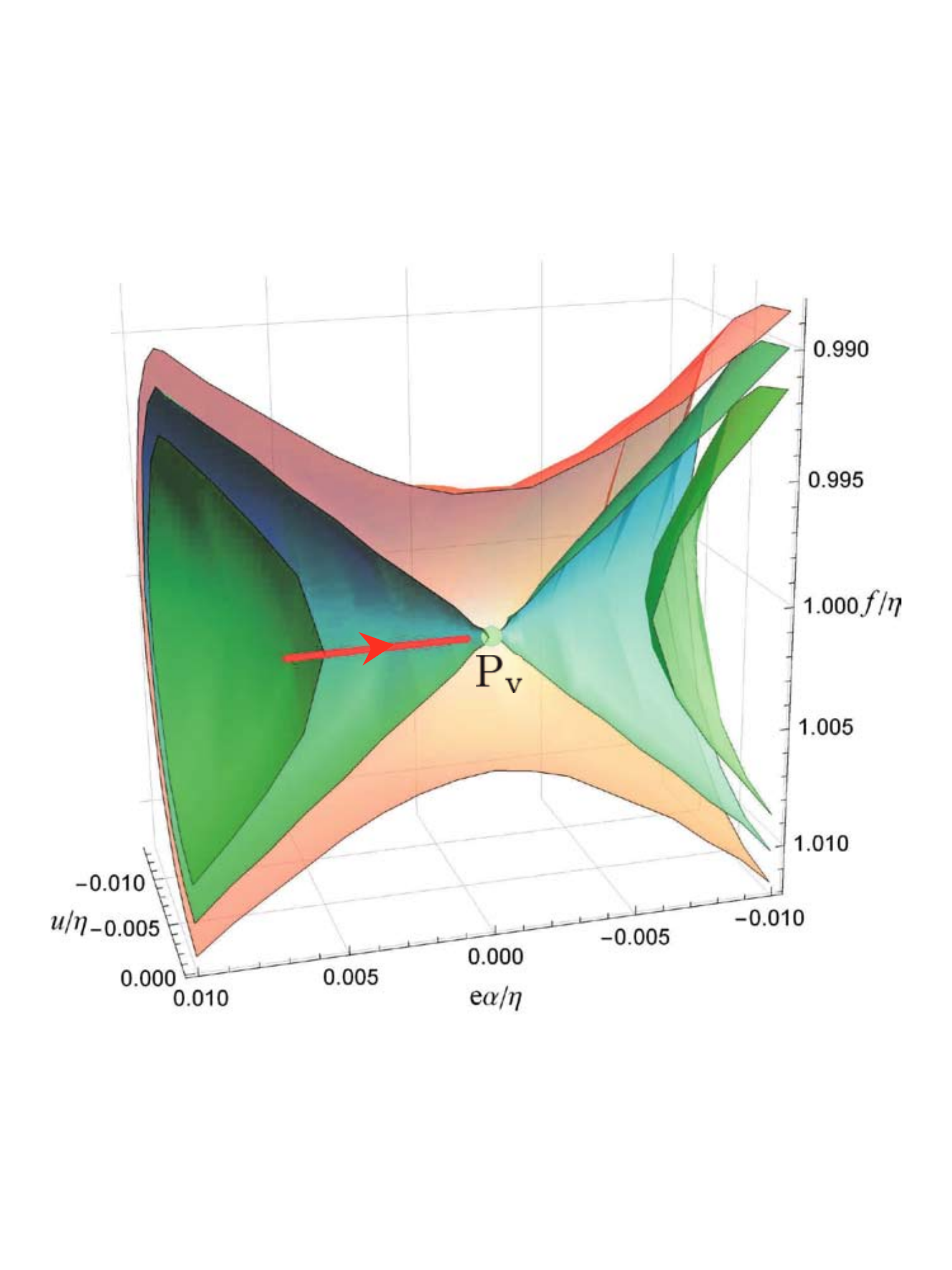}
\\
\vspace{-5mm}
(ii) ${\rm P_v} \to {\rm P_v} \ : \ (e=0.10 \ , \ \Omega=0.8109)$
\\
\includegraphics[width=5.5cm]{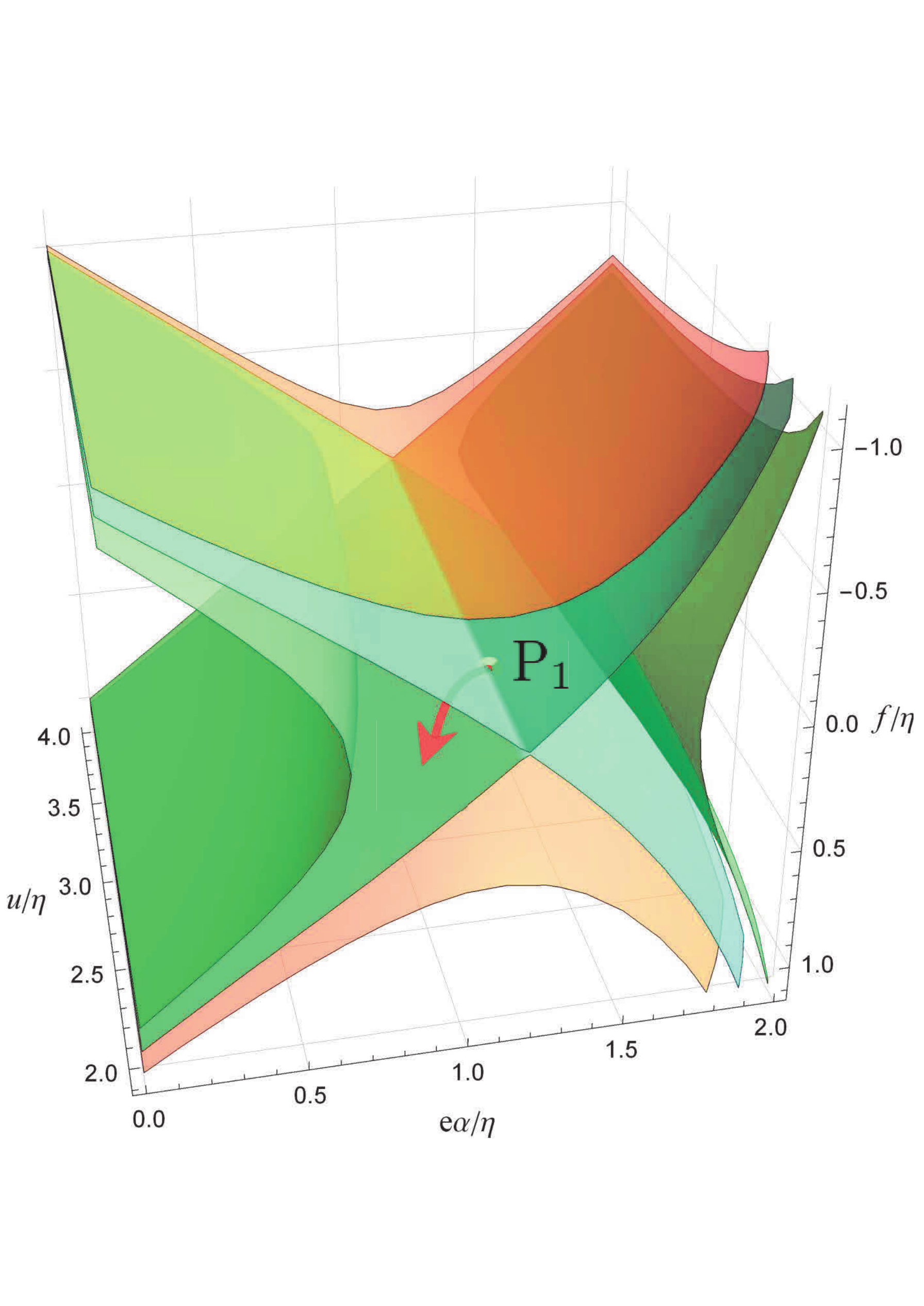}
\hspace{15mm}
\includegraphics[width=5.5cm]{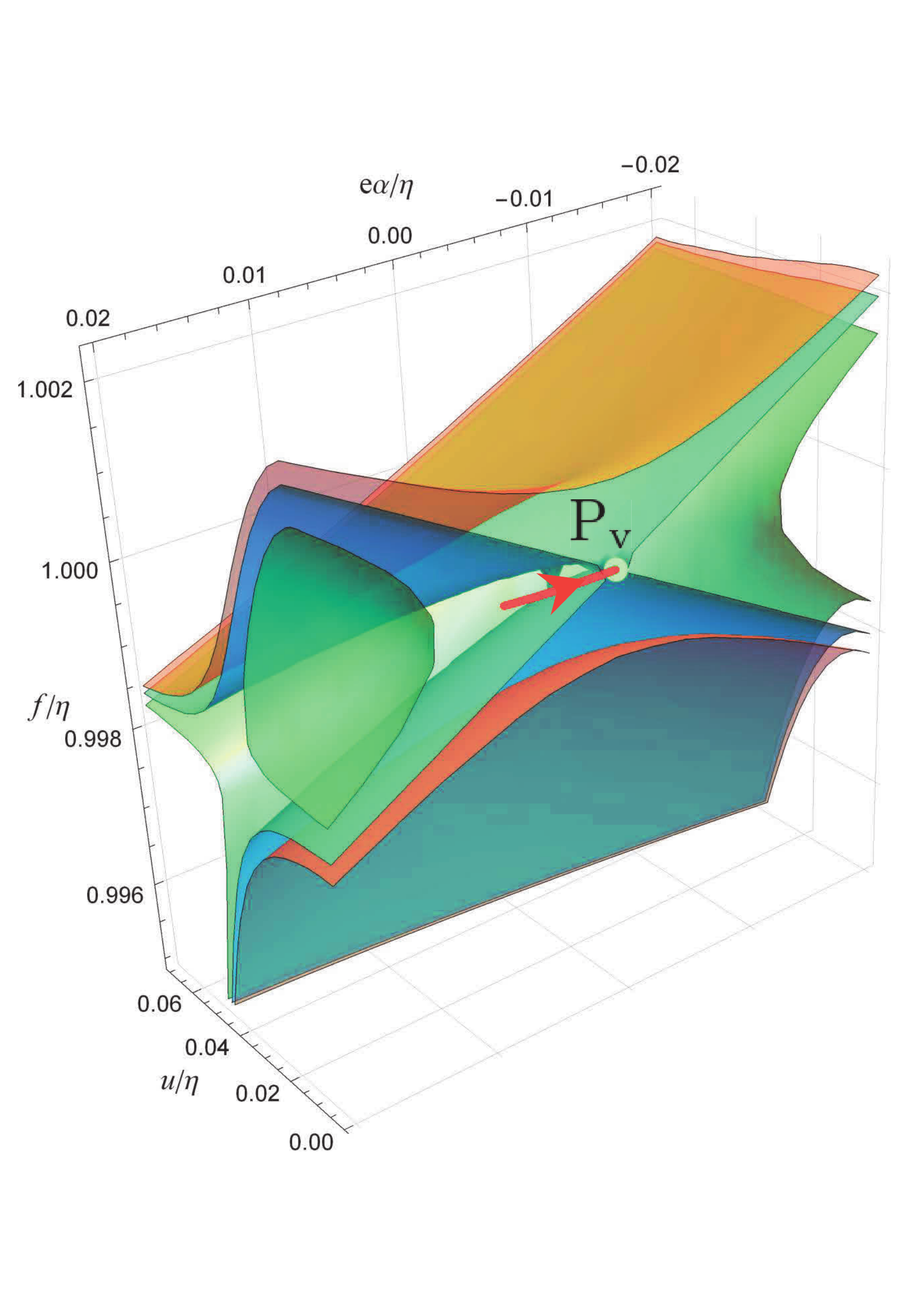}
\\
\vspace{-5mm}
(iii) ${\rm P_1} \to {\rm P_v} \ : \ (e=0.10 \ , \ \Omega=1.183)$
\caption{
Equisurfaces of the effective potential, $U_\text{eff}$, around the stationary points
in the three-dimensional space $(u, f, \alpha)$:
(i)
${\rm P_0}$ and ${\rm P_v}$,
(ii)
${\rm P_v}$ for both the start point and the end points,
(iii)
${\rm P_1}$ and ${\rm P_v}$.
The blue surface denote the values at the stationary points $U_\text{eff}({\rm P_0})$,
$U_\text{eff}({\rm P_1})$, and $U_\text{eff}({\rm P_v})$, respectively.
The orange surfaces denote larger, and the green surfaces denote smaller values
than the blue ones.
The motion of particle from and toward the stationary points are shown by arrows.
}
\label{fig:contour_bounce}
\end{figure}

At the stationary points, equisurfaces of the effective potential $U_{\rm eff}$
in the space $(u, f, \alpha)$ are depicted in Fig.\ref{fig:contour_bounce}.
In the same figure, segments of the trajectories
in the vicinity of the stationary points are shown.
The moving \lq particle\rq \ departs from the initial stationary point along an unstable direction,
and approaches to the terminal vacuum stationary point.

\subsection{The effective energy and the effective potential}

In this subsection, we discuss the properties of the effective energy, $E_\text{eff}$,
and the effective potential, $U_\text{eff}$, which characterize the numerical solutions
of the dynamical system \eqref{eq:S_eff}.
Physical energy given by the energy-momentum tensor $T_{\mu\nu}$ is discussed later
in the next section.

Inside the ball surfaces, $E_\text{eff}=U_\text{eff}=const.$ holds
because the functions $u,f,$ and $\alpha$ keep constant values.
The values of $U_\text{eff}$ at the stationary points are
$U_\text{eff}(\rm P_v)=0$, and $U_\text{eff}(\rm P_1)=-\lambda \eta^4/4$, and
$U_\text{eff}(\rm P_0)$, which depends on $\Omega$, approaches to $0$
as $\Omega\to \Omega_\text{min}$.
We see $E_\text{eff}=U_\text{eff}=0$ outside the ball surfaces in all cases
because $u,f,\alpha$ take the vacuum values.

\begin{figure}[H]
\vspace{1cm}
\centering
\includegraphics[width=5.3cm]{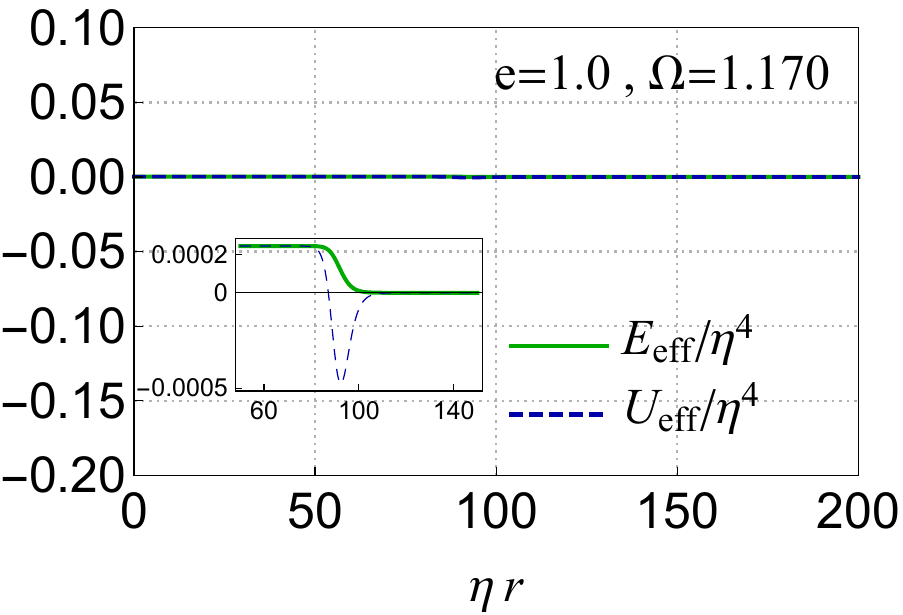}~~
\includegraphics[width=5.3cm]{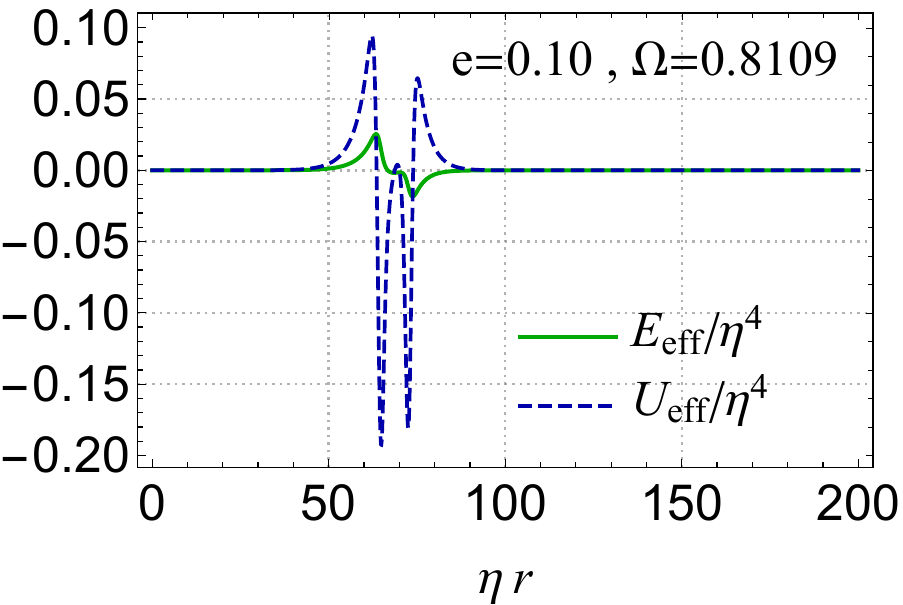}~~
\includegraphics[width=5.3cm]{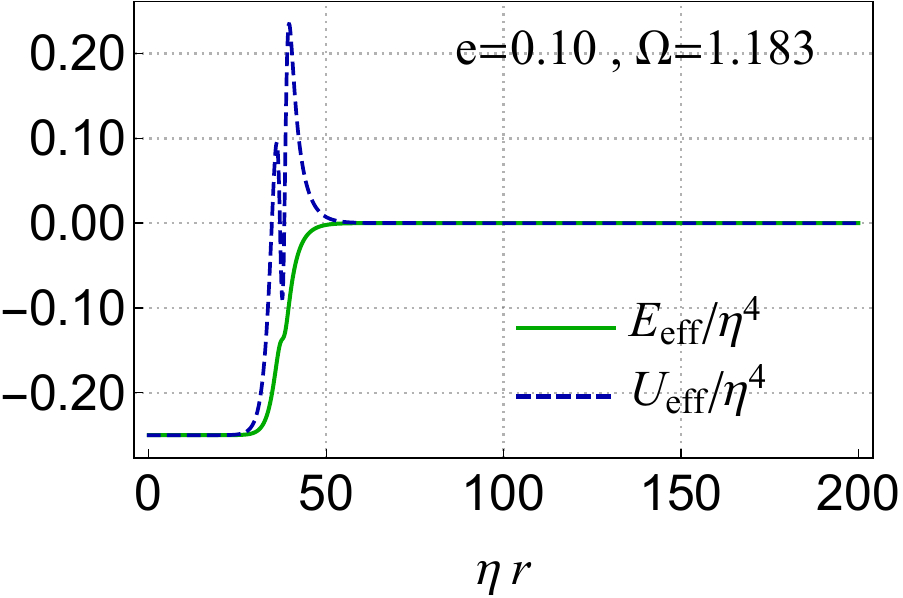}~~
\caption{
The values of the effective potential and the effective energy,
defined by \eqref{eq:U_eff} and \eqref{eq:E_eff}, are shown
for (0-V) type (left panel), (V-V) type (central panel), and (1-V) type (right panel)
as functions of $r$.
\label{fig:Effuctive_energy_potential}
}
\end{figure}

Fig.\ref{fig:Effuctive_energy_potential} represents the values of $E_\text{eff}$ and $U_\text{eff}$
of the solutions as functions of $r$.
In the case of (0-V) type (left panel), $E_\text{eff}$ and $U_\text{eff}$ keep almost zero
for all value of $r$. By magnification of the figure around the ball surface,
we see that $E_\text{eff}$ diminishes by a small amount at the ball surface.
This occurs owing to the work done by the friction forces activated around the ball surface.
The decrease of $E_\text{eff}$ caused by the friction forces
is equal to $U_\text{eff}(\rm P_0)-U_\text{eff}(\rm P_v)$.
In the limit $\Omega\to \Omega_\text{min}$, $U_\text{eff}(\rm P_0)\to U_\text{eff}(\rm P_v)=0$,
then, in such a solution, the friction forces become ineffective.
This actually occurs when the surface radius $r=R$ is large so that the friction
forces that are in proportion to $1/R$ are negligible.
Then, in the limit $\Omega\to \Omega_\text{min}$,
the radius of the NTS balls in the (0-V) type can be infinitely large.

In the case of (V-V) type (central panel), $U_\text{eff}$ oscillates
with a large amplitude in the shell region, and $E_\text{eff}$ does with a small amplitude.
In this type, the both initial and final stationary points are $\rm P_v$,
then the values of $U_\text{eff}$ inside and outside are exactly same.
One suspects that there would be no bounce solution that connects the same stationary
points under the existence of the friction force.
The numerical calculations show that $E_\text{eff}$ increases, decreases,
and increases again around the shell. This occurs by the friction forces of $u$ and $f$
and the anti-friction force of $\alpha$.
The works done by these forces compensate, then the initial value of $E_\text{eff}$ is recovered
during the evolution. Therefore, the bounce solutions that connect the same stationary
points $\rm P_v$ can exist in the system.
Keeping the cancellation, the friction and anti-friction forces,
whose magnitudes are in proportion to $1/R$, can become negligibly small
if the shell radius becomes large.
Then, the radius of the NTS balls in the (V-V) type can also be infinitely large.

In the case of (1-V) type (right panel), the difference of the effective potential
between the initial and final stationary points given by
\begin{align}
 	U_\text{eff}({\rm P_1})-U_\text{eff}({\rm P_v})=-\frac{\lambda}{4}\eta^4
\label{eq:effectiveU_potentialball}
\end{align}
is finite for finite $\lambda$ and $\eta$.
The effective energy rises up across the ball surface by a finite amount of
work done by anti-friction force of $\alpha$.
Thus, the radius of the (1-V) type NTS ball should be finite\footnote{
In the limit $\lambda\to 0$ or $\eta\to 0$, the maximum radius would be infinitely
large.} for the finite anti-friction force.

The solutions of the (0-V) type are found in a variety of field theories.
However, it should be noted that the solutions of the (V-V) and (1-V) types appear
for the system including the gauge field
whose kinetic term in the effective action \eqref{eq:S_eff} has the wrong sign
so that the equation of motion has the anti-friction term.

\newpage

\section{Internal properties of the NTS balls}
In this section, we study internal properties of the NTS balls for three types.
We explicitly show energy density, pressure, and charge densities of the solutions
of (0-V), (V-V), and (1-V) types.

\subsection{Energy density and pressure}

\begin{figure}[H]
\centering
\includegraphics[width=5.3cm]{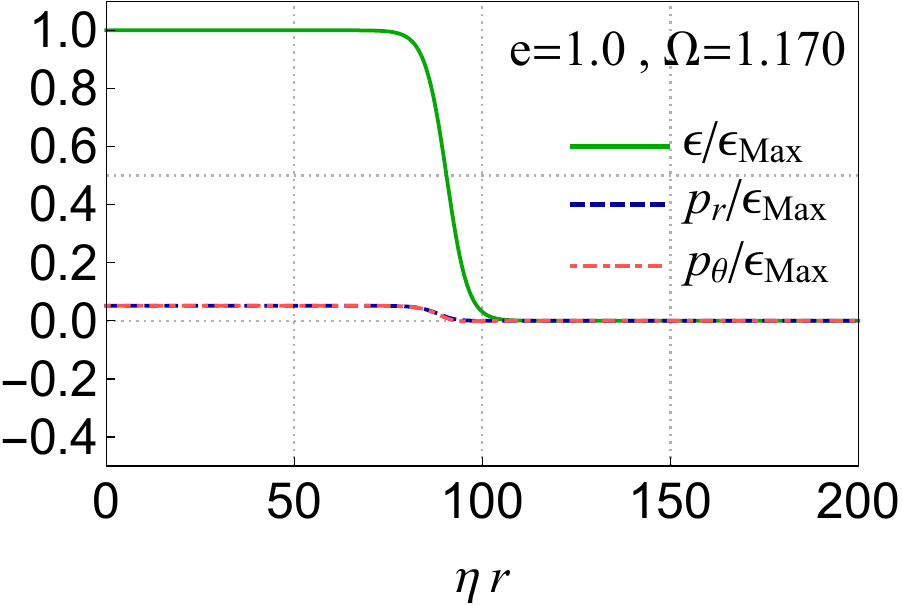}~~~~
\includegraphics[width=5.3cm]{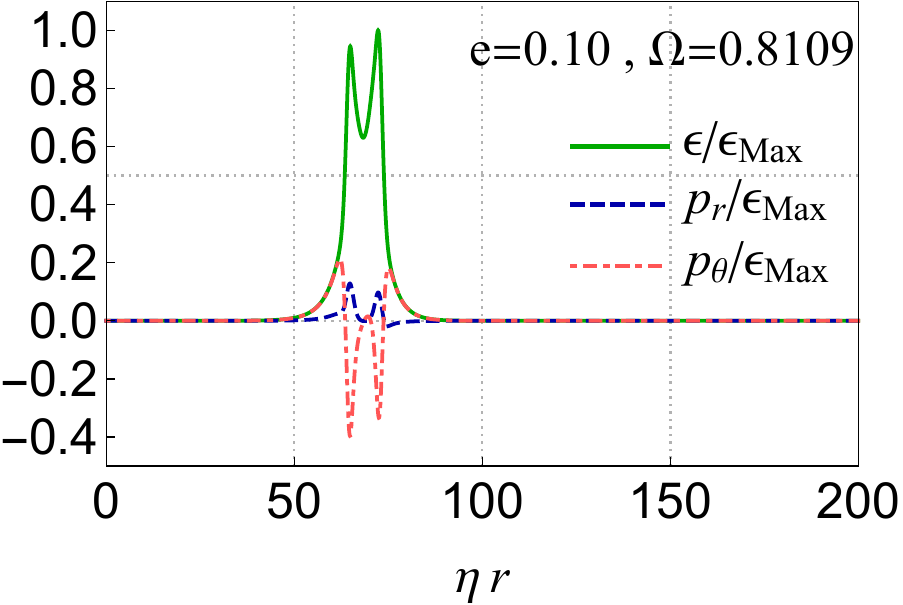}~~~~
\includegraphics[width=5.3cm]{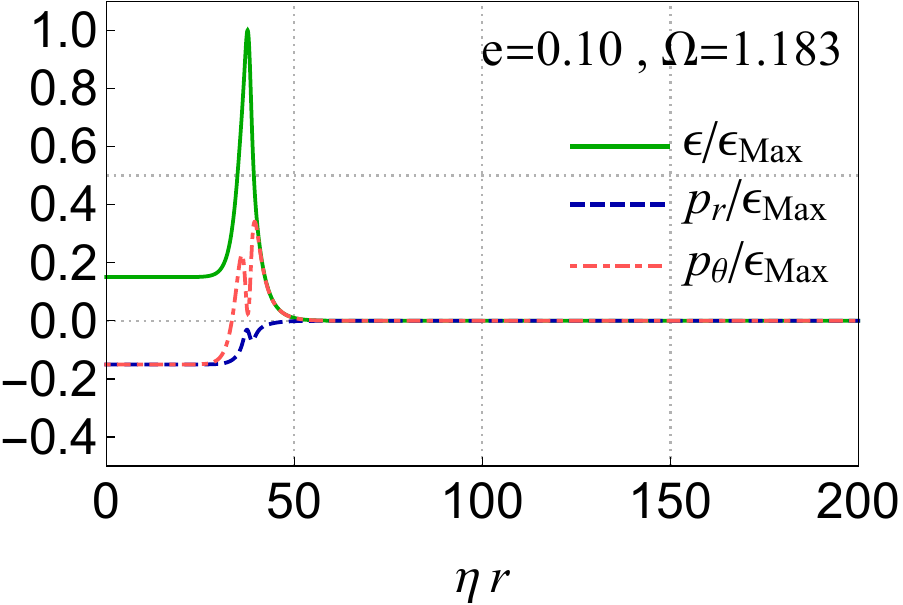}

\caption{
Energy density and pressure normalized by the maximum value of total energy density $\epsilon_\text{Max}$ are drawn
for (0-V) type (left panel), (V-V) type (central panel), and (1-V) type (right panel)
as functions of $r$.
\label{fig:energy_pressure}
}
\end{figure}

The energy density and the pressure inside the ball solutions are defined
by energy-momentum tensor $T_{\mu\nu}$,
whose expression is given in Appendix~\ref{E_M_tensor}.
Using the ansatz \eqref{eq:f}, \eqref{eq:u}, and \eqref{eq:alpha}, we see that
the energy density $\epsilon$, the radial pressure $p_r$, and the tangential pressure
$p_{\theta}(=p_{\varphi})$ of the system can be written by
\begin{align}
	&\epsilon = \epsilon_\text{$\psi$Kin}+\epsilon_\text{$\phi$Kin}
		+\epsilon_\text{$\psi$Elast}+\epsilon_\text{$\phi$Elast}
		+\epsilon_\text{Int} +\epsilon_\text{Pot}+\epsilon_\text{ES},
\label{eq:total_energy_density}
\\
&p_r = \epsilon_\text{$\psi$Kin}+\epsilon_\text{$\phi$Kin}
  +\epsilon_\text{$\psi$Elast}+\epsilon_\text{$\phi$Elast}
  -\epsilon_\text{Int} -\epsilon_\text{Pot}-\epsilon_\text{ES},
\label{eq:pressure_density_radius}
\\
&p_{\theta}= \epsilon_\text{$\psi$Kin}+\epsilon_\text{$\phi$Kin}
  -\epsilon_\text{$\psi$Elast}-\epsilon_\text{$\phi$Elast}
  -\epsilon_\text{Int} -\epsilon_\text{Pot}+\epsilon_\text{ES},
\label{eq:pressure_density_theta}
\end{align}
where
\begin{align}
	&\epsilon_\text{$\psi$Kin}:=\left|D_t\psi\right|^2
		=(e\alpha-\Omega)^2u^2, \quad
	\epsilon_\text{$\phi$Kin}:=\left|D_t\phi\right|^2=e^2f^2\alpha^2,
\cr
  	&\epsilon_\text{$\psi$Elast}:=\left(D_i\psi\right)^{\ast}\left(D^i\psi\right)=\biggl(\frac{du}{dr}\biggr)^2,\quad
	\epsilon_\text{$\phi$Elast}:=\left(D_i\phi\right)^{\ast}\left(D^i\phi\right)=\biggl(\frac{df}{dr}\biggr)^2, \quad
\cr
	&\epsilon_\text{Pot}:=V(\phi)=\frac{\lambda}{4}(f^2-\eta)^2, \quad
	\epsilon_\text{Int}:=\mu\left|\phi\right|^2\left|\psi\right|^2=\mu f^2u^2, \quad
	\epsilon_\text{ES}:=\frac{1}{2}E_iE^i=\frac{1}{2}\biggl(\frac{d\alpha}{dr}\biggr)^2,
\label{eq:energy_density}
\end{align}
are densities of the kinetic energy of $\psi$ and $\phi$, the elastic energy of $\psi$ and $\phi$,
the potential energy of $\phi$, the interaction energy between $\psi$ and $\phi$,
and the electrostatic energy, respectively.
In Fig.\ref{fig:energy_pressure} we show
$\epsilon$, $p_r$, and $p_{\theta}$ for three types of NTS ball solutions.

In the (0-V) type case, the energy density and the pressure can be represented approximately by
\begin{align}
  	\epsilon &\simeq \epsilon_\text{$\psi$Kin}+\epsilon_\text{$\phi$Kin}
			+\epsilon_\text{Int}+\epsilon_\text{Pot}
\cr
  	&=\frac{2}{\mu}e\alpha_0(\Omega-e\alpha_0)^3
		+\frac{1}{\mu}(e\alpha_0)^2(\Omega-e\alpha_0)^2
		+\frac{\lambda}{\mu^2}\left((\Omega-e\alpha_0)^2-\eta^2\right)^2,
\cr
	p_r &\simeq p_{\theta}
		\simeq \epsilon_\text{$\psi$Kin}+\epsilon_\text{$\phi$Kin}
		-\epsilon_\text{Int}-\epsilon_\text{Pot}
\cr
  		&=\frac{1}{\mu}(e\alpha_0)^2(\Omega-e\alpha_0)^2
		+\frac{\lambda}{\mu^2}\left((\Omega-e\alpha_0)^2-\eta^2\right)^2.
\end{align}
We see from Fig.\ref{fig:energy_pressure} that the pressure is considerably small
compare to the energy density for a (0-V) type solution.
Thus the equation of state is like non-relativistic fluid.
In the limit $\Omega\to \Omega_\text{min}$, the pressure of the ball vanishes,
namely, the ball consists of dust fluid \cite{Ishihara:2019gim}.
Therefore, we refer (0-V) type solutions as \textit{dust balls}.

For a (V-V) type solution, since the regions both inside and outside the ball
are in the vacuum state, $\epsilon=0$ and $p=0$,
and the energy density and the pressure are non-vanishing only in the shell region.
Then, we call (V-V) type solutions \textit{shell balls}.
The radial pressure, $p_r$, is rather small compare to the energy density
while the tangential pressure, $p_\theta$, takes negative non-negligible values,
i.e., the (V-V) type solution has a shell with tangential tension.

For a (1-V) type solution, using \eqref{eq:position_P1}, we find that
$\epsilon_\text{Pot}$, supplied by the potential of $\phi$, is the only nonvanishing component
inside the ball, where we see
\begin{align}
	\epsilon=-p_r=-p_{\theta}=-p_{\varphi}=\frac14 \lambda\eta^4.
  \label{eq:e_p_Pball}
\end{align}
Hence, we refer (1-V) type solutions as \textit{potential balls},
where a cosmological constant appears effectively inside the ball.
In the vicinity of the surface, a shell structure appears,
where the energy density and the pressure have some peaks.
In contrast to the shell balls, the tangential pressure is positive,
and the radial pressure is small as same as the shell balls.

\subsection{Charge density}
\begin{figure}[H]
\vspace{1cm}
\centering
\includegraphics[width=5.3cm]{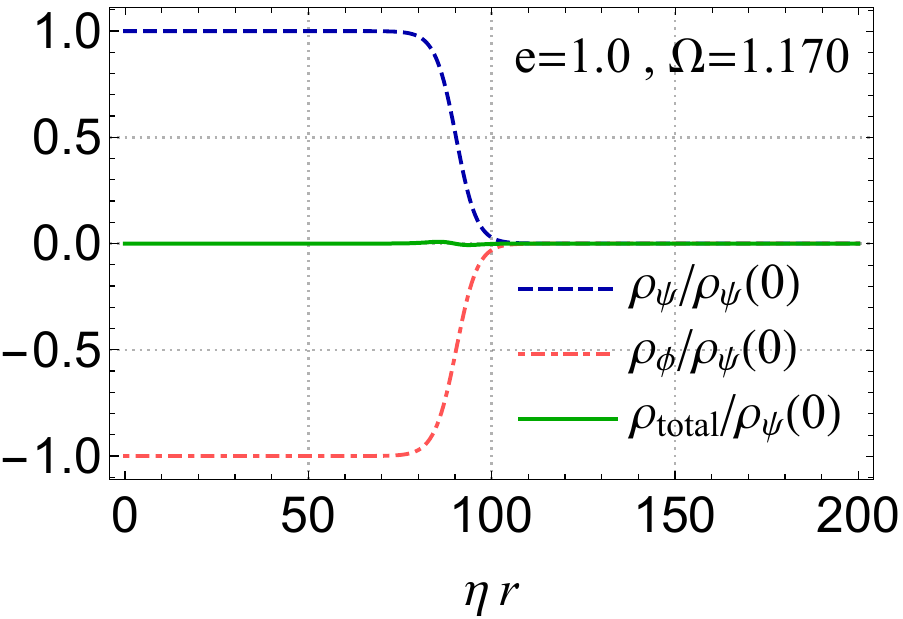}~~~~
\includegraphics[width=5.3cm]{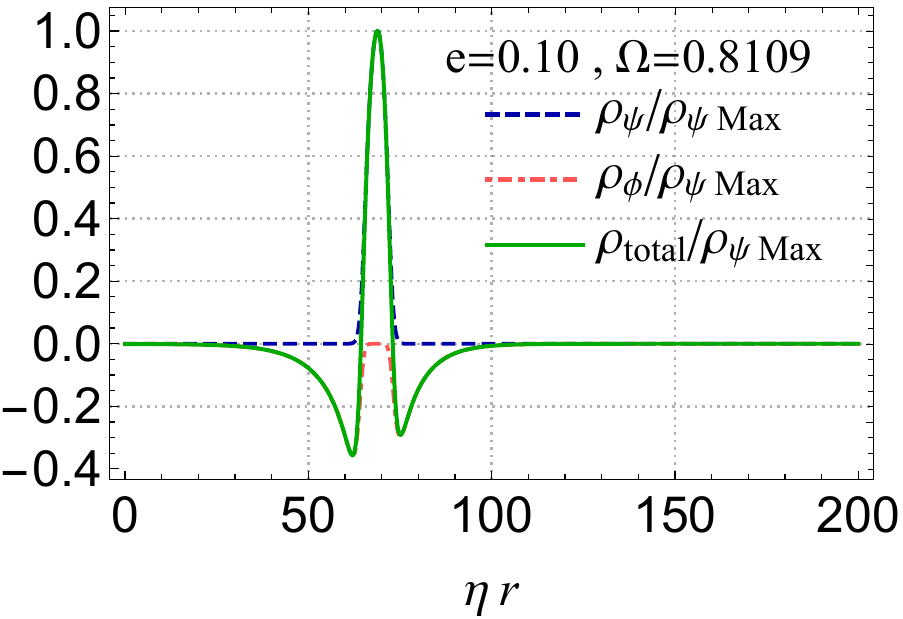}~~~~
\includegraphics[width=5.3cm]{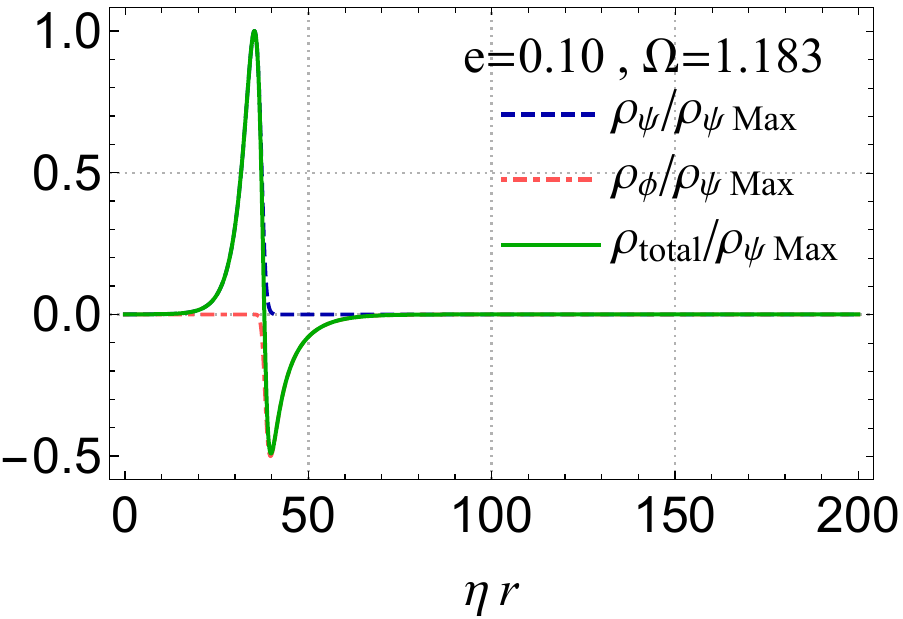}
\caption{
The charge densities of scalar fields, $\rho_{\psi}$, $\rho_{\phi}$,
and the total charge density $\rho_\text{total}$ normalized
by the maximum value of $\rho_{\psi}$ are shown
for (0-V) type (left panel), (V-V) type (central panel), and (1-V) type (right panel).
\label{fig:chargedensity}
}
\end{figure}

The charge densities $\rho_{\psi}$ and $\rho_{\phi}$, and the total charge density
$\rho_\text{total}=\rho_{\psi}+\rho_{\phi}$ are plotted in Fig.\ref{fig:chargedensity}
as functions of $r$.
In the case of the dust ball, the charge density $\rho_\psi$ is compensated by the counter charge
density $\rho_\phi$, then $\rho_\text{total}$ vanishes almost everywhere, namely,
perfectly screening occurs \cite{Ishihara:2018eah,Forgacs:2020sms}.
On the other hand, in the cases of the shell ball and the potential ball,
the charge densities are induced in the vicinity of the shell region.
For the potential balls, an electric double layer emerges at the surface,
while for the shell balls, an electric triple layer does (see Fig.\ref{fig:chargedensity}).
The total charge given by integration of the density is screened, i.e., $Q_{\psi}+Q_{\phi}=0$.
As a result, the NTS balls are observed as uncharged balls by a distant observer.

\newpage

\section{Mass, radius, and Stability of the NTS balls}

In this section, we inspect the total mass, $M_\text{NTS}$, and the total charge
of $\psi$, $Q_\psi$, for the numerical NTS-ball solutions in three types.
We discuss how $M_\text{NTS}$ and $Q_\psi$ depend on the ball radius $R$.

We define the total mass of the NTS-balls by
\begin{align}
	M_\text{NTS}=4\pi \int_0^{\infty} \epsilon(r)r^2 dr,
\label{eq:total_mass}
\end{align}
and the total charge of $\psi$ as
\begin{align}
	Q_\psi =4\pi \int_0^{\infty} \rho_\psi(r)r^2 dr.
\label{eq:total_charge}
\end{align}
Furthermore,
we define the ball radius $R$ for a numerical solution of NTS-ball by
\begin{align}
	4\pi \int_0^R \epsilon(r)r^2 dr = 0.99 M_\text{NTS},
\label{eq:radius}
\end{align}
so that 99 \% of the total mass of the NTS-ball is included within the radius $R$.

From Fig.\ref{fig:energy_pressure} and Fig.\ref{fig:chargedensity},
we see $\epsilon$ and $\rho_\psi$ are constant inside a dust ball,
then the both $M_\text{NTS}$ and $Q_\psi$ of the dust balls are proportional to their volume.
In contrast, for a shell ball, $\epsilon$ and $\rho_\psi$ concentrate on the shell region,
then the both $M_\text{NTS}$ and $Q_\psi$ are proportional to the surface area of
the shell balls.
For a potential ball, $\epsilon$ takes a constant value inside, and it
has a peak on the shell region, while $\rho_\psi$ is non-vanishing only in the shell region.
Then, $M_\text{NTS}$ depends on the surface area and the volume of the potential balls,
while $Q_\psi$ is proportional to only the surface area.
Therefore, $M_\text{NTS}$ and $Q_\psi$ depend on the radius $R$ as follows:
\begin{align}
\begin{tabular}{lll}
	$M_\text{NTS}\propto R^3$, \quad &$Q_{\psi}\propto  R^3$ & \quad \text{for dust balls,}
\\
	$M_\text{NTS}\propto R^2$, \quad &$Q_{\psi} \propto R^2$ & \quad {for shell balls,}
\\
	$M_\text{NTS}\sim \alpha R^2+\beta R^3$, \quad &$Q_{\psi}\propto R^2$ & \quad {for potential balls,}
\end{tabular}
\label{eq:totalCharge_radius_dependence}
\end{align}
where $\alpha$ and $\beta$ are some constants.

Defining the number of $\psi$ particles contained in a NTS-ball with $Q_{\psi}$ by
\begin{align}
	N_{\psi}:&=\frac{Q_{\psi}}{e},
\label{eq:psi_number}
\end{align}
we can define summation of the mass energy for $N_{\psi}$ free particles of $\psi$
that carry totally the same charge $Q_{\psi}$ of the NTS-ball by
\begin{align}
	M_\text{free} = N_{\psi} m_{\psi}= \frac{Q_{\psi}}{e} m_{\psi}.
\label{eq:energy_free}
\end{align}
If $M_\text{NTS}/M_\text{free}<1 $ for a NTS-ball, it has negative binding energy.
Then, the NTS-ball does not break up spontaneously into free particles, namely,
the NTS-ball is stable.

\begin{figure}[H]
\centering
\includegraphics[width=5.3cm]{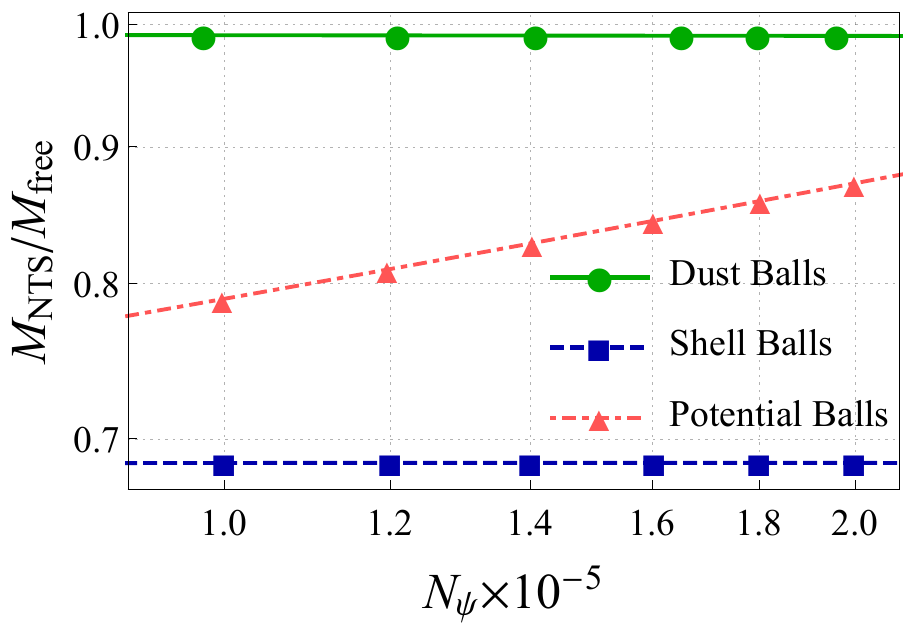}~~~~
\caption{
The mass ratio $M_\text{NTS}/M_\text{free}$ as functions of $N_{\psi}$ for three types of NTS-balls.
\label{fig:radius_energyratio}
}
\end{figure}

In Fig.\ref{fig:radius_energyratio}, we plot the ratio $M_\text{NTS}/M_\text{free}$
as a function of $N_{\psi}$.
The ratios $M_\text{NTS}/M_\text{free}$ for both the dust balls and the shell balls
are constant, since the both $M_\text{NTS}$ and $Q_\psi$ have the same dependence on the radius,
$\propto R^3$ for the dust balls, and $\propto R^2$ for the shell balls.
By the numerical calculations, we find $M_\text{NTS}/M_\text{free}$ are smaller than unity
for both the dust balls and the shell balls,
then these two types are stable without regard of their sizes.
In the case of the potential balls, $M_\text{NTS}$ and $Q_\psi$ depend differently
on $R$.
As shown in Fig.\ref{fig:radius_energyratio}, the potential balls are stable in the range
studied in this article,
however, it would be suspected that larger potential balls become unstable.

\newpage

\section{Summary}

We found numerically that three-types of stationary and spherically symmetric nontopological
soliton solutions, NTS-balls, with large size in the coupled system consisting of
a complex matter scalar field, a U(1) gauge field,
and a complex Higgs scalar field with a potential that causes spontaneous symmetry breaking.
By the assumption of the symmetries, the system is reduced to a dynamical system of
a particle described by an effective action with three-degrees of freedom where the radial
coordinate plays the role of the fictitious time.
The effective potential of the system has stationary points.
The vacuum stationary point, one of them, denotes the broken symmetry vacuum state.
There exist bounce solutions that connect one of the stationary point and the vacuum
stationary point describe NTS-balls with large size.

The effective action that describes the spherically symmetric system depends on the fictitious
time $r$ explicitly, then friction forces, in addition to the potential forces,
act on the moving particle.
The kinetic term of the gauge field has opposite sign to
the kinetic terms of the scalar fields.
Therefore, the friction forces work when the scalar fields change in their values,
while the anti-friction force works when the gauge field does.
This fact makes the system have a variety of solutions.

There exist three-types of solutions named dust balls, shell balls, and potential balls.
The dust ball, which appears in the case that the gauge coupling constant $e=1.0$,
has homogeneous energy density and negligibly small internal pressure inside the ball.
The shell ball, which appears in the case $e=0.10$, is a hollow sphere consisting vacuum region
surrounded by a shell with tangential tension.
The potential ball, which appears also in the case $e=0.10$, has a shell that encloses the region
filled by the potential energy of the Higgs scalar field.
The value of effective potential at the initial stationary point
is lower than the vacuum stationary point for the potential ball solutions.
Then, the anti-friction force that raises the effective energy plays the essential
role for the existence of the potential balls.

The friction force and anti-friction force are in proportion to the ball radius,
then in the cases that the effective potential at the initial stationary point can take
the same value at the vacuum stationary point, the cases of dust balls and shell balls,
the friction forces can be negligibly small,
namely the radius of the ball can be infinitely large.
In contrast, in the case that the effective potentials take different values at the initial and
vacuum stationary points, the case of potential balls, the friction and anti-friction forces should
yield work of the same amount of potential difference, then the ball radius should be finite.

All these NTS-balls have a common property, i.e.,
the charge of the complex matter scalar field is always totally screened
by the counter charge of the complex Higgs scalar field \cite{Ishihara:2018rxg,
Ishihara:2018eah,Ishihara:2019gim,Forgacs:2020vcy,Forgacs:2020sms}.
Then, in a viewpoint of a distant observer, all NTS-balls are electrically neutral objects.
It would be a desirable property as a dark matter candidate.
In order to apply such NTS-balls to the dark matter,
it is important next work to estimate how much amount of NTS-balls in the evolution of
the universe \cite{Frieman:1988ut,Griest:1989bq,Kasuya:2000wx,Postma:2001ea,Multamaki:2002hv,
Hiramatsu:2010dx}.

We showed that the mass of NTS-balls obtained  in this article are smaller than
the total mass of free particles condensed in the balls,
which is in proportion to the charge of the balls,
i.e., the NTS-balls have negative binding energy. This means that the NTS-balls do not
disperse into free particles.
For the dust balls and shell balls, the ratios of the mass and charge of the NTS-balls
are not depend on their radii,
while for the potential balls, the ratio is in proportion to the ball radius.
Then, the binding energy is negative for the dust balls and shell balls independently on their sizes,
while the potential balls could have positive binding energy if their sizes become large.
Various analysis of stability including perturbative analisys \cite{Cohen:1986ct,Kusenko:1997ad,
Multamaki:1999an,Paccetti:2001uh,Kawasaki:2005xc,Sakai:2007ft} are also important future works.

Study of the gravitational fields of the NTS-balls is an interesting and important
issue \cite{Friedberg:1986tp,Friedberg:1986tq,Lee:1986tr,Lynn:1988rb,Mielke:2002bp}.
The geometries outside the NTS-balls are described by the Schwarzschild metrics because
the regions are commonly spherically symmetric vacua.
In contrast, internal geometry of the NTS-balls depends on the internal properties for three-types.
Then, observation of the geometry inside the balls distinguishes the type of solutions.
Furthermore, it is also interesting whether the NTS-balls can be exotic compact objects,
namely NTS-balls could have ISCOs and photon spheres and so on.
A relativistic compact NTS-ball, if it could be possible,
would be expected as an alternative or a seed of a black hole.

\section*{Acknowledgements}

We would like to thank K.-i. Nakao, H. Itoyama, Y.Yasui, N. Maru, N. Sakai, and M. Minamitsuji
for valuable discussion.
H.I. was supported by JSPS KAKENHI Grant Number 16K05358.
\newpage

\appendix

\section{Linear Analysis at the Stationary Points}
\label{stability}

We linearize the equations of motion \eqref{eq:eq_u}, \eqref{eq:eq_f} and \eqref{eq:eq_alpha}
at the stationary points in the form
\begin{align}
  \frac{d^2X}{dr^2}+\frac{2}{r}\frac{d X}{dr}+AX=0,
  \label{eq:linear_equation}
\end{align}
where $X$ denotes deviation from a stationary point as
\begin{align}
  X:=
  \begin{pmatrix}
    u- u_{\rm st} \\
    f- f_{\rm st} \\
    \alpha- \alpha_{\rm st}
  \end{pmatrix} \ , \
\end{align}
where $( u_{\rm st}, f_{\rm st}, \alpha_{\rm st})$ denotes the position of one of
the stationary points listed in \eqref{eq:position_Pv}-\eqref{eq:position_P2}.
The matrix $A$ is given by
\begin{align}
  A:=
  \begin{pmatrix}
  (e \alpha_{\rm st}-\Omega)^2-\mu f_{\rm st}^2 & -2\mu f_{\rm st} u_{\rm st} & 2e u_{\rm st}(e \alpha_{\rm st}-\Omega)
   \\
   -2\mu f_{\rm st} u_{\rm st} & e^2 \alpha_{\rm st}^2-\frac{\lambda}{2}(3 f_{\rm st}^2-\eta^2)
	-\mu u_{\rm st}^2 & 2e^2 f_{\rm st} \alpha_{\rm st}
 \\
   -4e u_{\rm st}(e \alpha_{\rm st}-\Omega) & -4e^2 f_{\rm st} \alpha_{\rm st} & -2e^2( f_{\rm st}^2+ u_{\rm st}^2)
  \end{pmatrix}.
\label{matrix_A}
\end{align}
At large $r$, the second term, friction term,
in \eqref{eq:linear_equation} can be negligible.
If the matrix $A$ has a real negative eigenvalue or a complex eigenvalue,
there exists an unstable linear solution of $X$ that grows or decays exponentially.
The decaying solutions are necessary for the bounce solutions.
In this appendix, we show that there exist unstable directions of the equations of
motion at the stationary points, P$_V$, P$_0$, and P$_1$ on $R_1$, while no unstable direction
at any point on $R_2$.

\subsubsection{The vacuum stationary point ${\rm P_v}$}
Substituting $( u_{\rm st}, f_{\rm st}, \alpha_{\rm st})=( u_{\rm v}, f_{\rm v}, \alpha_{\rm v})$
into \eqref{matrix_A}, we get
\begin{align}
  A=
  \begin{pmatrix}
   -\mu\eta^2+\Omega^2 & 0 & 0
   \\
   0 & -\lambda\eta^2 & 0
   \\
  0 & 0 & -2e^2\eta^2
  \end{pmatrix}.
  \label{eq:matrix_Pv}
\end{align}
Two eigenvalues are negative, and the rest is also negative
if $\Omega_{\rm max}^2:=\mu\eta^2>\Omega^2$.
In the case $\Omega_{\rm max}^2>\Omega^2$, all fields can decay exponentially toward
the vacuum stationary point P$_V$.

\subsubsection{The stationary point ${\rm P_0}$}
For large NTS-ball solutions, where the friction terms in \eqref{eq:S_eff} are negligible,
The effective energy defined by \eqref{eq:E_eff} is conserved for the (0-V)-type solutions.
Then, we require $U_{\rm eff}(\rm P_0) \simeq U_{\rm eff}(\rm P_v)$.
The equality holds for $\Omega = \Omega_{\rm min}$,
which is defined in \eqref{eq:Omega_limit_min}.
Then, we inspect the eigenvalues of $A$ for $\Omega= \Omega_{\rm min}$.

The characteristic equation of $A$ is in the form
\begin{align}
	F(\kappa):=\kappa^3 + \alpha \kappa^2 +\beta\kappa +\gamma=0,
\label{characteristic_eq}
\end{align}
where $\kappa$ is an eigenvalue of $A$, $\alpha, \beta$ are complicated functions of $f_0, u_0$,
and $\alpha_0$ and $\gamma$ is explicitly given by
\begin{align}
 \gamma:=32 e^2f_0^2 u_0^2(-2 e^2\alpha_0^2 +8e \alpha_0 f_0
	+(3 \lambda-2 ) f_0^2 - \lambda\eta^2 ).
  \label{eq:eigenfunction2}
\end{align}
It can be shown that if $\mu>\lambda$,
\begin{align}
  F(0)=\gamma=64\frac{e^2}{\mu}\left(\frac{\lambda}{\mu}\right)^{3/2}
	\left(\frac{1-\sqrt{\lambda/\mu}}{\sqrt{2-\lambda/\mu}}\right)^2 \eta^6 >0,
\end{align}
then $A$ has a real negative eigenvalue, at least.

\subsubsection{The stationary point $\rm P_1$ on the ridge ${\rm R_1}$}
Setting $( u_{\rm st}, f_{\rm st}, \alpha_{\rm st})=( u_1, f_1, \alpha_1)$,
we obtain
\begin{align}
  A=
  \begin{pmatrix}
   0 & 0 & 0
   \\
    0 & \Omega^2+\frac{\lambda}{2}\eta^2-\mu u_1^2 & 0
   \\
  0 & 0 & -2e^2 u_1^2
  \end{pmatrix},
  \label{eq:matrix_L1}
\end{align}
where $u_1$ is an arbitrary constant.
One of the eigenvalues is zero because the stationary point $\rm P_1$ is on a ridge ${\rm R_1}$.
If $u_1\neq 0$, one eigenvalue is negative at least,
and if $\mu u_1^2>\Omega^2+\frac{\lambda}{2}\eta^2$,
the rest of eigenvalue is also negative.

\subsubsection{The stationary point $\rm P_2$ on the ridge ${\rm R_2}$}
Setting $( u_{\rm st}, f_{\rm st}, \alpha_{\rm st})=( u_2, f_2, \alpha_2)$,
the matrix $A$ is reduced to
\begin{align}
  A
  =
  \begin{pmatrix}
   (e \alpha_2-\Omega)^2 & 0 & 0
   \\
    0 & e^2 \alpha_2^2+\frac{\lambda}{2}\eta^2 & 0
   \\
  0 & 0 & 0
  \end{pmatrix},
  \label{eq:matrix_L2}
\end{align}
where $\alpha_2$ is an arbitrary constant.
In this case, one eigenvalue is zero because $\rm P_2$ is on the ridge
${\rm R_2}$, and the rest two are non-negative.
Thus, there is no unstable direction at the stationary point $\rm P_2$.

\section{Energy-Momentum Tensor of the System}
\label{E_M_tensor}
The energy-momentum tensor $T_{\mu\nu}$ of the present system is given by
\begin{align}
T_{\mu\nu}
	=&2(D_{\mu}\psi)^{\ast}(D_{\nu}\psi)
	-g_{\mu\nu}
		(D_{\alpha}\psi)^{\ast}(D^{\alpha}\psi)
\cr
	&+2(D_{\mu}\phi)^{\ast}(D_{\nu}\phi)
	-g_{\mu\nu}(D_{\alpha}\phi)^{\ast}(D^{\alpha}\phi)
\cr
	&-g_{\mu\nu}\left( V(\phi)
	 + \mu \psi^{\ast}\psi \phi^{\ast} \phi \right)
\cr
	&+\left( F_{\mu\alpha}F_{\nu}^{~\alpha}
	-\frac{1}{4}g_{\mu\nu}F_{\alpha\beta}F^{\alpha\beta}\right).
\label{eq:T_munu}
\end{align}
Energy density and pressure components are given by
\begin{align}
 	\epsilon=&-T_t^t
\cr
		= &\left|D_{t}\psi\right|^2  +(D_{i}\psi)^{\ast}(D^{i}\psi)
		 +\left|D_{t}\phi\right|^2+(D_{i}\phi)^{\ast}(D^{i}\phi)
\cr
	& +V(\phi)+\mu|\psi|^2|\phi|^2 +\frac{1}{2}\left(E_iE^i+B_iB^i\right),
\label{eq:T_tt}
\end{align}
\vspace{-15mm}
\begin{align}
 	p_r=&T_r^r
\cr		= &(D_r\psi)^{\ast}(D^r \psi)+\left|D_{t}\psi\right|^2
			-(D_{\theta}\psi)^{\ast}(D^{\theta}\psi) -(D_{\varphi}\psi)^{\ast}(D^{\varphi}\psi)
\cr
		 &+(D_r\phi)^{\ast}(D^r \phi)+\left|D_{t}\phi\right|^2
			-(D_{\theta}\phi)^{\ast}(D^{\theta}\phi)-(D_{\varphi}\phi)^{\ast}(D^{\varphi}\phi)
\cr
	  	& -V(\phi) -\mu|\psi|^2|\phi|^2
\cr
		&+\frac{1}{2}(-E_rE^r+E_\theta E^\theta+E_\varphi E^\varphi
		-B_r B^r+B_\theta B^\theta+B_\varphi B^\varphi),
\label{eq:T_rr}
\end{align}
\vspace{-15mm}
\begin{align}
 	p_{\theta}=&T_\theta^\theta
\cr		= &(D_\theta\psi)^{\ast}(D^\theta \psi)
		+\left|D_{t}\psi\right|^2 -(D_r\psi)^{\ast}(D^r\psi)
		-(D_{\varphi}\psi)^{\ast}(D^{\varphi}\psi)
\cr
		 &+(D_{\theta}\phi)^{\ast}(D^{\theta}\phi)+\left|D_{t}\phi\right|^2
		-(D_r\phi)^{\ast}(D^r \phi) -(D_{\varphi}\phi)^{\ast}(D^{\varphi}\phi)
\cr
	  	& -V(\phi) -\mu|\psi|^2|\phi|^2
\cr
		&+\frac{1}{2}(-E_\theta E^\theta +E_rE^r+E_\varphi E^\varphi
		-B_\theta B^\theta+B_r B^r+B_\varphi B^\varphi),
\label{eq:T_thetatheta}
\end{align}
\vspace{-15mm}
\begin{align}
 	p_{\varphi}=&T_\varphi^\varphi
\cr		= &(D_{\varphi}\psi)^{\ast}(D^{\varphi}\psi)
		+\left|D_{t}\psi\right|^2 -(D_r\psi)^{\ast}(D^r\psi)
		-(D_\theta\psi)^{\ast}(D^\theta \psi)
\cr
		 &+(D_{\varphi}\phi)^{\ast}(D^{\varphi}\phi)+\left|D_{t}\phi\right|^2
		-(D_r\phi)^{\ast}(D^r \phi) -(D_{\theta}\phi)^{\ast}(D^{\theta}\phi)
\cr
	  	& -V(\phi) -\mu|\psi|^2|\phi|^2
\cr
		&+\frac{1}{2}(-E_\varphi E^\varphi+E_rE^r+E_\theta E^\theta
		-B_\varphi B^\varphi+B_r B^r+B_\theta B^\theta).
\label{eq:T_phiphi}
\end{align}

\section{Estimation of the maximum radius of the potential balls}
\label{potential_radius}
In the case of the potential balls, (1-V)-type, we have
\begin{align}
  U_\text{eff}({\rm P_v})=0 \ , \  U_\text{eff}({\rm P_1})=-\frac{\lambda}{4}\eta^4.
\label{eq:E_potential}
\end{align}
Therefore, the difference of $U_\text{eff}$ must be compensated by the work defined
by \eqref{eq:E_fri}.
In this section, we estimate the radius of the potential balls by calculate the
work done by the friction forces.

We roughly estimate the work \eqref{eq:E_fri} by
\begin{align}
	W\sim  \frac{2}{R}\left(\left(\frac{\Omega/2e}{2r_u}\right)^2 2r_u
	+\left(\frac{\Omega/2e}{2r_A}\right)^2 2r_A\right)
	-\frac{4}{R}\left(\left(\frac{\eta}{4r_{\phi}}\right)^2 4r_{\phi}
	+\left(\frac{u_1}{4r_{\psi}}\right)^2 4r_{\psi}\right),
  \label{eq:E_fri3}
\end{align}
where $r_{\psi}:=m_{\psi}^{-1}=(\sqrt{\mu}\eta)^{-1},
r_{\phi}:=m_{\phi}^{-1}=(\sqrt{\lambda}\eta)^{-1}$ and
$r_A:=m_A^{-1}=(\sqrt{2}e\eta)^{-1}$
are Compton lengths of the gauge field,
the complex matter scalar field, and the complex Higgs scalar field, respectively.

Requiring $W=\lambda\eta^4/4$,
we obtain that the radius of the potential balls as
\begin{align}
	R\sim \frac{2\sqrt{2}\Omega^2}{\lambda e \eta^3}\left(\frac{u_1+\eta}{2\eta}\right)
	-\frac{4e}{\lambda e \eta^3}\left(\sqrt{\lambda}\eta^2+\sqrt{\mu}u_1^2\right).
\label{eq:radius_3}
\end{align}
The radius obtained above depends on the parameters of the system
and $u_1$, the central value of the function $u(r)$.
Numerically, $u_1$ is the same order of magnitude of the symmetry breaking scale $\eta$.
Then, we can estimate the radius as
\begin{align}
	R\sim \frac{2\sqrt{2}\Omega^2}{\lambda e \eta^3}
	-\frac{4e}{\lambda e \eta^3}\left(\sqrt{\lambda}\eta^2
	+\sqrt{\mu}\eta^2\right)
	=4\Omega^2 (r^2_{\phi}r_A)-4r^2_{\phi}\left(\frac{1}{r_{\phi}}
	+\frac{1}{r_{\psi}}\right).
\label{eq:radius_4}
\end{align}
This estimation is in accord with the numerical calculations that show
the radius of the potential balls increases as the parameter $\Omega$ increases.
As mentioned in Appendix A, $\Omega < \Omega_{\rm max}$
in order that the function $u(r)$ decays exponentially toward P$_V$.
Using $\Omega_\text{max}$, we see that the maximum radius for the potential balls is given by
\begin{align}
	R_\text{max}\sim 4 r_{\phi}\left(\frac{r_{\phi}r_A}{r^2_{\psi}}
	+\frac{r_{\phi}}{r_{\psi}}-1\right).
\label{eq:radius_Max}
\end{align}


\end{document}